\documentclass[
 reprint,
 superscriptaddress,
 amsmath,amssymb,
 aps,
floatfix,
]{revtex4-2}

\usepackage{graphicx}
\usepackage{dcolumn}
\usepackage{bm}
\usepackage{siunitx}
\begin{document}

\preprint{APS/123-QED}

\title{Brownian motion of droplets induced by thermal noise}

\author{Haodong Zhang}
\author{Fei Wang}%
\email{fei.wang@kit.edu}
\affiliation{Institute of Applied Materials-Microstructure Modelling and Simulation, Karlsruhe Institute of Technology (KIT), Strasse am Forum 7, Karlsruhe 76131, Germany}
\affiliation{Institute of Nanotechnology, Karlsruhe Institute of Technology (KIT), Hermann-von-Helmholtz-Platz 1, Eggenstein-Leopoldshafen 76344, Germany}

\author{Lorenz Ratke}%
\affiliation{Institute of Materials Research, German Aerospace Center, Linder Hoehe, 51147 Cologne, Germany}

\author{Britta Nestler}%
\affiliation{Institute of Applied Materials-Microstructure Modelling and Simulation, Karlsruhe Institute of Technology (KIT), Strasse am Forum 7, Karlsruhe 76131, Germany}
\affiliation{Institute of Digital Materials Science, Karlsruhe University of Applied Sciences, Moltkestrasse 30, Karlsruhe 76133, Germany}

\date{\today}

\begin{abstract}
\textcolor{black}{Brownian motion (BM) is pivotal in natural science for the stochastic motion of microscopic droplets. 
In this study, we investigate BM driven by thermal composition noise at sub-micro scales, 
where inter-molecular diffusion and surface tension both are significant.
To address BM of microscopic droplets, 
we develop two stochastic multi-phase-field models coupled with the full Navier-Stokes equation,
namely Allen-Cahn-Navier-Stokes (ACNS) and Cahn-Hilliard-Navier-Stokes (CHNS).
Both models are validated against capillary wave theory;
the Einstein's relation for the Brownian coefficient $D^*\sim k_B T/r$ at thermodynamic equilibrium is recovered. 
Moreover, by adjusting the co-action of the diffusion, Marangoni effect, and viscous friction, 
two non-equilibrium phenomena are observed. 
(I) The droplet motion transits from the Brownian to Ballistic with increasing Marangoni effect which is emanated from the energy dissipation mechanism distinct from the conventional fluctuation-dissipation theorem.
(II) The deterministic droplet motion is triggered by the noise induced non-uniform velocity field which leads to a novel droplet coalescence mechanism associated with the thermal noise.
}
\end{abstract}

\maketitle

\section{Introduction}
\textcolor{black}{Brownian motion (BM) owes its original observation and nomenclature to Robert Brown~\cite{brown1828brief}.
Within the purview of natural sciences, BM characterizes the persistent stochastic motion of particles suspended within a fluid medium~\cite{kramers1940brownian,wang1945theory}. 
Given its interdisciplinary significance and broad accessibility, BM has not only elicited considerable scientific interest but also found widespread utility in numerous practical applications~\cite{saffman1975brownian,shukla2016effective}.
A salient feature of the particle BM is the relationship of the Brownian coefficient $D^*\sim k_B T/(\eta\, r_0)$ with the viscosity $\eta$ and particle radius $r_{\scriptscriptstyle 0}$.
Resulting from this relationship, the particle mean squared displacement (MSD) increases linearly with time as 
\begin{equation}
    \langle\Delta X^{\scriptscriptstyle 2}\rangle\sim 2\,D^* t.
\end{equation}
It is important to note that this relation is based on certain fundamental assumptions. 
Firstly, it presupposes a rigid particle without fluid flow inside the particle.
Secondly, it does not account for the hydrodynamic effects arising from the surface tension, 
which are particularly pertinent when changes in droplet curvature due to deformation and coalescence are pronounced. 
Lastly, the mutual interactions between particles are neglected and only the viscous friction with the surrounding matrix is considered.
However, for lots of materials, including but not limited to polymer solutions, aerogels, and hydrogels~\cite{wang2016gelation,baudron2019continuous}, 
the particles do not exhibit rigid, solid-body behavior; instead, they manifest as liquid-state entities, 
often taking the form of droplets via phase separation or nucleation. 
These droplets assume a paramount role in shaping the microstructural composition of materials, 
for which the mass diffusion, surface tension effect~\cite{mohraz2016interfacial}, and hydrodynamics ~\cite{zaccone2013colloidal}  
are pivotal. 
Furthermore, the dispersed droplets engage in interactions 
and coagulation with one another, 
rendering the complex process inherently contextualized 
within a many-body physical framework,
which is beyond the scope of the single droplet BM discussed by Einstein~\cite{einstein1905molekularkinetischen}.}

\textcolor{black}{
A well-established model for characterizing BM is the Langevin equation~\cite{tothova2011langevin}.
Within this framework, BM is depicted as the motion of a particle represented as a mass point, 
propelled by the stochastic force resulting from collisions between the matrix fluid molecules and the Brownian particles. 
Meanwhile, this motion is counteracted by the viscous friction force, as described by Stokes' formula.
Incorporating specific force terms into the Langevin equation allows for the investigation of diverse 
and complex BM scenarios across various domains. 
Examples include BM coupled with a magnetic field~\cite{lucero2020brownian}, BM influenced by laser-induced temperature field~\cite{hajizadeh2017brownian}, and BM subjected to electric fields~\cite{tothova2020brownian}.
Furthermore, the discourse on Brownian motion extends to the realm of anisotropic particles, where researchers often employ an orientation-dependent Langevin model~\cite{mayer2021two,lin2023noise}.
Expanding the scope, the inclusion of hydrodynamic memory effects~\cite{alder1970decay}, 
which account for the inertial influence of the surrounding matrix on Brownian particles, 
has led to the study of non-Markovian BM.
This is accomplished by introducing a history-dependent 
Basset–Boussinesq–Oseen force term into the Langevin model~\cite{seyler2019long}.
Each of these instances adds distinctive features to the conventional Brownian motion model. 
The intricate interplay between complex fields and thermal fluctuations leads to an apparent transition in particle dynamics. 
At short time intervals, these transitions manifest as a shift from Brownian behavior, 
with $\langle\Delta X^2\rangle\sim t$, to ballistic behavior, with $\langle\Delta X^2\rangle\sim t^2$, 
yet retaining Brownian behavior at longer time scales.
Notably, the resultant motion closely resembles the self-propulsion observed in biological entities, such as E. coli bacteria and spermatozoa~\cite{fier2018langevin}.
This is of paramount importance in comprehending 
the dynamics of active matter, 
a phenomenon often referred as active Brownian motion (ABM). 
Consequently, corresponding fluctuation-dissipation theorem (FDT) is also tailored to ABM, 
grounded in the principles of Langevin mechanics~\cite{burkholder2019fluctuation}. }

\textcolor{black}{
Confucius once espoused the wisdom of 
revisiting established theories to uncover new knowledge. 
In this research endeavor, we revisit the passive BM of sub-micrometer droplets 
and focus on the thermal composition noises. 
These noises emerge from the fluctuations in chemical free energy, a factor not typically considered in traditional Langevin mechanics.
A notable scenario is reported in Ref.~\cite{klosin2020phase}, 
cellular protein droplets exhibit pronounced composition noise~\cite{shin2017liquid},
while demonstrating weak to moderate hydrodynamic effects 
due to the dynamic arrest of proteins~\cite{elbaum2015disordered,shin2017liquid}. 
Given the significance of protein molecule stability under thermal noise 
as well as the importance of Brownian motion-induced protein droplet coalescence~\cite{lee2023size} for live organisms, 
we have developed two distinct phase-field models. 
These models aim to explore the Brownian motion of tiny droplets in the presence of intense composition noise 
coupled with varying degrees of hydrodynamics.}

\textcolor{black}{In our model, we establish an equivalence between Langevin mechanics and the Navier-Stokes equation to consider the fluid dynamics, 
as outlined in~\cite{pomeau2017langevin}. 
To account for the crucial role of composition noise in tiny droplets, 
we introduce the stochastic phase-field equation, 
which imparts substantial significance to random behaviors.
While the composition fluctuations is predominantly dissipated by diffusion, 
it remains intricately linked to the strong surface tension effect 
which converts perturbations in chemical free energy into fluctuations in kinetic energy, 
and results in the so-called Marangoni flow. 
This transformative mechanism falls beyond the purview of the conventional fluctuation-dissipation theorem 
and Langevin mechanics.
Furthermore, employing a multicomponent multiphase model, we explore the interactions of multiple droplets, incorporating the interplay of diffusion, surface tension-induced Marangoni effects, and viscous friction. 
The inclusion of composition noise not only introduces a new degree of freedom to equilibrium Brownian motion 
but also precipitates various non-equilibrium behaviors.
}

\textcolor{black}{This paper is structured as follows: 
in Sec.~\ref{sec:model}, two mathematical models are presented. The numerical stability
of the model discretization
is validated with the capillary wave theory in Sec.~\ref{sec:cwt}. 
In Sec.~\ref{sec:equi}, we study the equilibrium of the droplet motion with both models,
where the Einstein's theory $D^*\sim k_B T /r_0$ is recovered. 
In Sec.~\ref{sec:nonequi}, 
we show two non-equilibrium behaviors stemming from the thermal noise. 
Firstly, the single droplet motion transforms from Brownian to Ballistic which indicates a distinct energy dissipation mechanism from FDT.
Secondly, the multi-droplet simulation shows an underlying coalescence mechanism that is generated by the noise induced non-uniform velocity field. 
A comparison with previous Golovin-Tanaka coalescence mechanism is made;
differences from previous work are discussed.}

\section{Numerical model}\label{sec:model}
\subsection{Stochastic Allen-Cahn-Navier-Stokes model}
\label{sec:acns_model}
We consider a system consisting of $N$ Brownian particles ($N\in \mathbb{Z}$) 
in a domain $\Omega$.
The Brownian particles are characterized by a vector of
phase order parameters $\boldsymbol{\phi}(\boldsymbol{x},t)$ in the Gibbs simplex 
$G=\{ \boldsymbol{\phi}\in \mathbb{R}^{\scriptscriptstyle N+1}:\sum_{\alpha=0}^{\scriptscriptstyle N}\phi_{\scriptscriptstyle \alpha}=1,
\ \phi_{\scriptscriptstyle \alpha}\geq0 \}$. The index $0$ is especially assigned for the matrix with $\phi_{\scriptscriptstyle 0}:=\phi_{\scriptscriptstyle M}$.
The chemical free energy functional $\mathcal{F}$ of the system is written as a function of the order parameter as 
\begin{equation}
 \mathcal{F} = \int_{\scriptscriptstyle\Omega} \bigg[g(\boldsymbol{\phi}) 
 + \epsilon\, a(\boldsymbol{\phi},\nabla\boldsymbol{\phi})
 +\frac{\omega(\boldsymbol{\phi})}{\epsilon}\bigg] d\Omega.
 \label{eq:F_ac} 
\end{equation}
Here, $\epsilon$ is a parameter related to the interface width.
The bulk chemical free energy density is defined by $g(\boldsymbol{\phi})$, which 
is used to ensure volume conservation acting as the role of a
Lagrangian multiplier.
Similar to the work of Landau~\cite{landau2013course}, the gradient energy density $a(\boldsymbol{\phi},\nabla\boldsymbol{\phi})$ 
is formulated by a generalized asymmetric expression as~\cite{steinbach1996phase}  
\begin{equation}
 a(\boldsymbol{\phi},\nabla\boldsymbol{\phi}) = \sum_{\scriptscriptstyle\alpha<\beta}^{\scriptscriptstyle N,N}
 \gamma_{\scriptscriptstyle\alpha\beta}(\phi_{\scriptscriptstyle\alpha}\nabla\phi_{\scriptscriptstyle\beta}
 -\nabla\phi_{\scriptscriptstyle\alpha}\phi_{\scriptscriptstyle\beta})^{\scriptscriptstyle 2},
 \label{eq:a}
\end{equation}
where $\gamma_{\scriptscriptstyle\alpha\beta}$ denotes the interfacial tension between $\alpha$ and $\beta$ phases. 
The last chemical free energy contribution $\omega(\boldsymbol{\phi})$ is formulated as a non-convex function with $N+1$ global minima, 
which characterize the equilibrium states of the $N$ particles and the matrix. 
To save computational effort, 
we adopt a multi-obstacle potential~\cite{wang2015phase,wu2019droplets,cai2021phase} as
\begin{equation}
  \!\!\!\omega(\boldsymbol{\phi}) \!=\! \left\{
    \begin{array}{ll}
      \!\displaystyle\frac{16}{\pi^{\scriptscriptstyle 2}}\sum\limits_{\scriptscriptstyle \alpha<\beta}^{\scriptscriptstyle N, N}
     \!\!\gamma_{\scriptscriptstyle \alpha\beta}\phi_{\scriptscriptstyle\alpha}\phi_{\scriptscriptstyle\beta}
     +\!\!\!\sum\limits_{\scriptscriptstyle \alpha<\beta<\gamma}^{\scriptscriptstyle N, N, N}\!\!\!\!\chi^{\scriptscriptstyle *}
     \phi_{\scriptscriptstyle\alpha}\phi_{\scriptscriptstyle\beta}\phi_{\scriptscriptstyle\gamma}, & \text{if } \boldsymbol{\phi}\text{ in } G, \\[2pt]
      +\infty, & \text{else}.
  \end{array} \right.\label{eq:omega}
\end{equation}
Here, the penalty parameter $\chi^{\scriptscriptstyle *}$ models the triple interactions in the system
and avoids the occurrence of the third particle contribution at two particle interfaces. 
Physically, 
the choice of $\chi^{\scriptscriptstyle *}$
affects the evolution of the energy landscape 
from the bulk of one particle to another bulk following the 
gradient descent trajectory;
the value of $\chi^{\scriptscriptstyle *}$
reflects the triple molecular interaction in a lattice model and 
should be determined by the phase diagram.

Under the assumption of incompressibility, 
the evolution of the phase-field variable $\boldsymbol{\phi}$ and the fluid velocity $\boldsymbol{u}$
in the system is controlled by the Allen-Cahn-Navier-Stokes (ACNS) equations~\cite{hohenberg1977theory} as
\begin{gather}
  \nabla\cdot \boldsymbol{u} = 0,\\[2pt]
(\partial_{\scriptscriptstyle t}\boldsymbol{\phi} + \boldsymbol{u} \cdot \nabla \boldsymbol{\phi}) = -\frac{1}{P\acute{e}}\Big(\tau 
 \boldsymbol{\mu}_{\scriptscriptstyle \phi}
  + \sqrt{\tau}\,\boldsymbol{\xi}_{\scriptscriptstyle\phi} + \lambda\Big),\label{eq:AC}\\[3pt]
   \rho\,(\partial_{\scriptscriptstyle t} \boldsymbol{u} \!+\! \boldsymbol{u} \!\cdot \!\nabla \boldsymbol{u})
  \!= \!\frac{-\nabla P\!-\!\boldsymbol{\phi}\nabla\boldsymbol{\mu}_{\scriptscriptstyle \phi}}{We} \!+ \!\frac{\eta\nabla^{\scriptscriptstyle 2}\boldsymbol{u}}{Re} \!+\!  \frac{\sqrt{\eta}\,\boldsymbol{F}}{Re}
\label{eq:NS1}.
\end{gather}
The kinetic parameter $\tau$ 
controls the evolution rate of $\boldsymbol{\phi}$ towards equilibrium. 
The chemical potential $\boldsymbol{\mu}_{\scriptscriptstyle \phi}$ is defined as $\delta \mathcal{F}/\delta \boldsymbol{\phi}$. 
The Lagrange multiplier $\lambda$ ensures the constraint $\sum_{\scriptscriptstyle \alpha=0}^{\scriptscriptstyle N}\phi_{\scriptscriptstyle \alpha}=1$ by taking the following formulation
\begin{align}
\lambda = \frac{1}{N+1}\sum_{\scriptscriptstyle \alpha=1}^{\scriptscriptstyle N+1}\big(\tau\,\mu_{\scriptscriptstyle\phi_\alpha}+\sqrt{\tau}\,\xi_{\scriptscriptstyle\phi_\alpha}\big).
\end{align}
The pressure is labeled as $P$. 
The density $\rho$ and the dynamic viscosity $\eta$ are linearly 
interpolated over the individual densities and viscosities of each particle
and matrix as $\rho=\sum_{\scriptscriptstyle \alpha=0}^{\scriptscriptstyle N}\rho_{\scriptscriptstyle \alpha}\phi_{\scriptscriptstyle \alpha}$ and $\eta=\sum_{\scriptscriptstyle \alpha=0}^{\scriptscriptstyle N}\eta_{\scriptscriptstyle \alpha}\phi_{\scriptscriptstyle \alpha}$, respectively. In the discussion section on the ACNS model, the subscripts $_P$ and $_M$
stand for the proprieties of the particle and the matrix, respectively.
Without losing the generality, the density and the viscosity of the matrix are set as the reference value to vary the corresponding values of the particles in the current work.
The dimensionless quantities P\text{$\acute{e}$}, Re, and We are calculated by the characteristic velocity $u^{\scriptscriptstyle *}$, diffusivity $D^{\scriptscriptstyle *}$, length $x^{\scriptscriptstyle *}$, density $\rho^{\scriptscriptstyle *}$, viscosity $\eta^{\scriptscriptstyle *}$ and surface tension $\sigma^{\scriptscriptstyle *}$ as 
\begin{equation*}
     P\acute{e}=\frac{u^* x^*}{D^*}, \quad Re=\frac{\rho^* u^* x^*}{\eta^*}, \quad We=\frac{\rho^* u^{*2} x^*}{\sigma^*}.
\end{equation*}
In this work, we set Re$=1.0$, P\text{$\acute{e}$}$=1.0$, and We$=100$, if not specified differently. 

Noteworthily, two types of stochastic processes are coupled in the ACNS model, namely, (a) the composition fluctuation $\boldsymbol{\xi}_{\scriptscriptstyle\phi}=(\xi_{\scriptscriptstyle\phi_0}, \xi_{\scriptscriptstyle\phi_2},\dots,\xi_{\scriptscriptstyle\phi_{N}})=(\xi_{\scriptscriptstyle\phi}, \xi_{\scriptscriptstyle\phi},\dots,\xi_{\scriptscriptstyle\phi})$, and (b) the random body force term $\boldsymbol{F}$. Both noises are
 Gaussian and spatial/temporal relevant as
\begin{gather}
  \big\langle \xi_{\scriptscriptstyle\phi}
  ,  
  \xi_{\scriptscriptstyle\phi}^{\scriptscriptstyle\prime}
  \big\rangle
  =\frac{2\, k_{\scriptscriptstyle B}T}{v_{\scriptscriptstyle l}\,\Delta t } \delta(\boldsymbol x-\boldsymbol x^{\scriptscriptstyle\prime})\delta(t-t^{\scriptscriptstyle\prime}),\label{eq:noise_phi}\\[3pt]
   \big\langle \boldsymbol{F}
   , \boldsymbol{F}^{\scriptscriptstyle\prime}
   \big\rangle
   =\frac{2\, k_{\scriptscriptstyle B}T}{v_{\scriptscriptstyle l}\, \Delta t}\nabla^{\scriptscriptstyle 2} \delta(\boldsymbol x-\boldsymbol x^{\scriptscriptstyle\prime})\delta(t-t^{\scriptscriptstyle\prime}),\label{eq:noise_force}
\end{gather}
According to the fluctuation-dissipation theorem (FDT), 
the noise amplitudes are decided by the Boltzmann constant $k_{\scriptscriptstyle B}$, the temperature $T$, the lattice volume $v_{\scriptscriptstyle l}$, and the simulation time step $\Delta t$.

\subsection{Stochastic Cahn-Hilliard-Navier-Stokes model}
We postulate another conserved Cahn-Hilliard-Navier -Stokes model to characterize Brownian particles 
in a domain $\Omega$ by its
composition $c(\boldsymbol{x},t)$, 
while the matrix composition is $1-c(\boldsymbol{x},t)$.
The chemical free energy functional $\mathcal{F}$ of the system is then written as 
\begin{equation}
 \mathcal{F} = \int_{\scriptscriptstyle\Omega} \Big[
 f(c) + \sigma\epsilon\big(\nabla c\big)^{\scriptscriptstyle 2}\Big] d\Omega.
 \label{eq:F_ch} 
\end{equation}
Here, the liquid-matrix interfacial tension $\gamma$ is determined by the surface tension parameter $\sigma$~\cite{wang2012spinodal}. 
The parameter $\epsilon$ is related to the interface width. 
The bulk chemical free energy density $f(c)$ takes the regular solution formulation as~\cite{zhang2021phase}
\begin{equation}
 f = \frac{R_{\scriptscriptstyle g}T}{v_{\scriptscriptstyle m}}\Big[c\ln c + (1-c)\ln (1-c)\Big]+\chi\, c\,(1-c),
 \label{eq:f} 
\end{equation} 
where $R_{\scriptscriptstyle g}$ denotes the universal gas constant, 
$v_{\scriptscriptstyle m}$ represents the molar volume. 
The molecular interaction between droplet and matrix is scaled by the Flory parameter $\chi$ 
which takes $3.78$ for 
the system with an upper critical point phase diagram in this work.
Under the assumption of incompressibility, 
the evolution of the Brownian particle composition $c$ and the fluid velocity $\boldsymbol{u}$
in the system is governed by the stochastic Cahn-Hilliard-Navier-Stokes (CHNS) equations~\cite{hohenberg1977theory} as
\begin{gather}
  \nabla\cdot \boldsymbol{u} = 0,\notag\\[2pt]
 \partial_{\scriptscriptstyle t}c+ \boldsymbol{u} \cdot \nabla c =  \frac{\nabla}{P\acute{e}}\cdot\big(\mathcal{M}\nabla \mu + \sqrt{\mathcal{M}}\,\boldsymbol\xi_{\scriptscriptstyle c}\big),\label{eq:CH}\\[3pt]
 \rho\,(\partial_{\scriptscriptstyle t} \boldsymbol{u}\!+\! \boldsymbol{u} \!\cdot\! \nabla \boldsymbol{u})
  \!=\! \frac{-\nabla P\!-\!c\nabla\mu}{We} \! + \!\frac{\eta\nabla^{\scriptscriptstyle 2}\boldsymbol{u}}{Re}\!+\!\frac{\sqrt{\eta}\,\boldsymbol{F}}{Re} 
   \label{eq:NS2}.
\end{gather}
Here, the
chemical potential $\mu=\delta\mathcal{F}/\delta c=\partial f/\partial c - 2\sigma\epsilon\nabla^{\scriptscriptstyle 2} c$ is magnified by the mobility $\mathcal{M}$, which propels the diffusion. 
The mobility takes Onsager's relation as $\mathcal{M}=\mathcal{M}_{\scriptscriptstyle 0}c (1-c)$ with $\mathcal{M}_0=(\epsilon/\sigma)(D_{\scriptscriptstyle P}(1-c) + D_{\scriptscriptstyle M}c)$. 
$D_{\scriptscriptstyle P}$ and  $D_{\scriptscriptstyle M}$ stand for the self-diffusivity of the droplet and matrix, respectively~\cite{zhang2022janus}.
The surface tension force $-c\,\nabla\mu$ induces the convection
and the pressure term $P$ is solved by the Poisson equation resulting from the incompressible flow. 
The density $\rho$ and the dynamic viscosity $\eta$ are linearly 
interpolated as 
$\rho(c)=(\rho_1-\rho_2)c+\rho_2$, 
$\eta=(\eta_1-\eta_2)c+\eta_2$.
At the droplet (matrix) equilibrium composition $c_{\scriptscriptstyle 0}$ ($c_{\scriptscriptstyle 1}$), we have $\rho(c_{\scriptscriptstyle 0})=\rho_{\scriptscriptstyle P}$, $\eta(c_{\scriptscriptstyle 0})=\eta_{\scriptscriptstyle P}$ ($\rho(c_{\scriptscriptstyle 1})=\rho_{\scriptscriptstyle M}$, $\eta(c_{\scriptscriptstyle 1})=\eta_{\scriptscriptstyle M}$).
Alike the previous ACNS model, two types of stochastic processes are coupled in the CHNS model, namely, (i) the composition fluctuation flux term $\boldsymbol\xi_{\scriptscriptstyle c}$ as 
\begin{align*}
   \boldsymbol\xi_{\scriptscriptstyle c} 
   =\begin{cases}
    (\xi_{\scriptscriptstyle x})=(\xi_{\scriptscriptstyle c}),  &\quad\text{for 1D}; \\[2pt]
    (\xi_{\scriptscriptstyle x},\xi_{\scriptscriptstyle y})=(\xi_{\scriptscriptstyle c},\xi_{\scriptscriptstyle c}),  &\quad\text{for 2D}; \\[2pt]
    (\xi_{\scriptscriptstyle x},\xi_{\scriptscriptstyle y},\xi_{\scriptscriptstyle z})=(\xi_{\scriptscriptstyle c},\xi_{\scriptscriptstyle c},\xi_{\scriptscriptstyle c}),  &\quad\text{for 3D},
  \end{cases}
\end{align*}
whose component is Gaussian and spatial/temporal relevant as 
\begin{gather}
  \big\langle \xi_{\scriptscriptstyle c}
  , \xi_{\scriptscriptstyle c}^{\scriptscriptstyle\prime}
  \big\rangle
  = \frac{2\,k_{\scriptscriptstyle B}T}{v_{\scriptscriptstyle l}\,\Delta t}\nabla^{\scriptscriptstyle 2}\delta(\boldsymbol x \!-\!\boldsymbol x^{\scriptscriptstyle\prime})\delta(t\!-\!t^{\scriptscriptstyle\prime}),
  \label{eq:noise_c}
\end{gather}
For example, the composition noise amplitude of pure water at $298.15\si{K}$ takes $0.203$ after non-dimensionalization (see Supplementary).
(ii) $\boldsymbol{F}$ is identical to Eq.~\eqref{eq:noise_force} of the ACNS model.

\subsection{Simulation setup and boundary conditions}
The finite difference method is implemented
on a staggered mesh with a size of $Nx\times Ny\times Nz$ and equidistant Cartesian spacing $\Delta x=\Delta y=\Delta z$ to solve the evolution equations, namely, the Allen-Cahn equation Eq.~\eqref{eq:AC} and the Cahn-Hilliard equation Eq.~\eqref{eq:CH}.
The Navier-Stokes equations, Eq.~\eqref{eq:NS1} and Eq.~\eqref{eq:NS2} are updated with the explicit Euler scheme, the phase-field variables $\boldsymbol{\phi}$, the concentration $c$, and the fluid velocity $\boldsymbol{u}$ are subjected to 
the periodic boundary conditions. 
The discretized space and time steps are shown in Tab.~SI in the supplementary. 
Parallelization of the numerical algorithm is achieved with Message Passing Interface (MPI) techniques. 
The numerical convergence of the model is demonstrated in the next section. 
The simulations are performed on the parallel computer bwUniCluster of Baden-Württemberg equipped with Intel Xeon Gold CPUs in the environment of Red Hat Enterprise.

\section{Validation}\label{sec:cwt}
\subsection{Capillary wave theory}
For liquid surfaces with small thermal noise, 
the capillary wave theory (CWT) is regarded as a decent way 
to describe its behavior which has been proven by several experiments~\cite{fisher1982agreement,davis1977capillary}
and simulations~\cite{pushkarev2000turbulence}. 
Perturbed by the noise, 
the surface energy increase $\Delta E$ of a planar fluid interface is proportional to the surface area change as 
\begin{align}
    \Delta E 
    \approx \frac{\gamma}{2} \int \big(
    \nabla h\big)^{\scriptscriptstyle 2} \,dx dy, \label{eq:de}
\end{align}
where the interface position $h$ is defined by the location 
with composition $\phi_{\scriptscriptstyle P}=0.5$ for ACNS, 
and $c=0.5$ for the CHNS model.
The liquid-matrix interfacial tension is represented by $\gamma$.
After Fourier transformation, 
Eq.~\eqref{eq:de} is expressed in the reciprocal space as 
\begin{align*}
    \Delta E(q) =\frac{\gamma}{2} \int 
    q^{\scriptscriptstyle 2}\big\rvert \Delta\widetilde{h}(q)\big\rvert^{\scriptscriptstyle 2} d q,
\end{align*}
where $q$ symbolizes the wave number, 
and $\widetilde{h}(q)$ represents
the capillary wave amplitude in the wavenumber domain. 
In statistical mechanics, 
each wave mode of the fluctuation has the identical energy $k_{\scriptscriptstyle B} T$, 
so that
\begin{align}
    \big\langle \Delta\widetilde{h}^{\scriptscriptstyle 2}(q)\big\rangle
    = \frac{k_{\scriptscriptstyle B} T}{4\pi^{\scriptscriptstyle 2}q^{\scriptscriptstyle 2}\,\gamma
    },
    \label{eq:cwt_q}
\end{align}
where $\big\langle \Delta\widetilde{h}^{\scriptscriptstyle 2}(q)\big\rangle$ is named as the structure factor of the fluid interface. 
According to the noise formulations stated by Eq.~\eqref{eq:noise_phi} and Eq.~\eqref{eq:noise_c}, 
we have $\big\langle \Delta\widetilde{h}^{\scriptscriptstyle 2}(q)\big\rangle \propto \xi_{\scriptscriptstyle \phi}^{\scriptscriptstyle 2}$ 
and $\big\langle \Delta\widetilde{h}^{\scriptscriptstyle 2}(q)\big\rangle \propto\xi_{\scriptscriptstyle c}^{\scriptscriptstyle 2}$ for ACNS and CHNS, respectively.

\begin{figure*}[htbp]
 \includegraphics[width=0.86\linewidth]{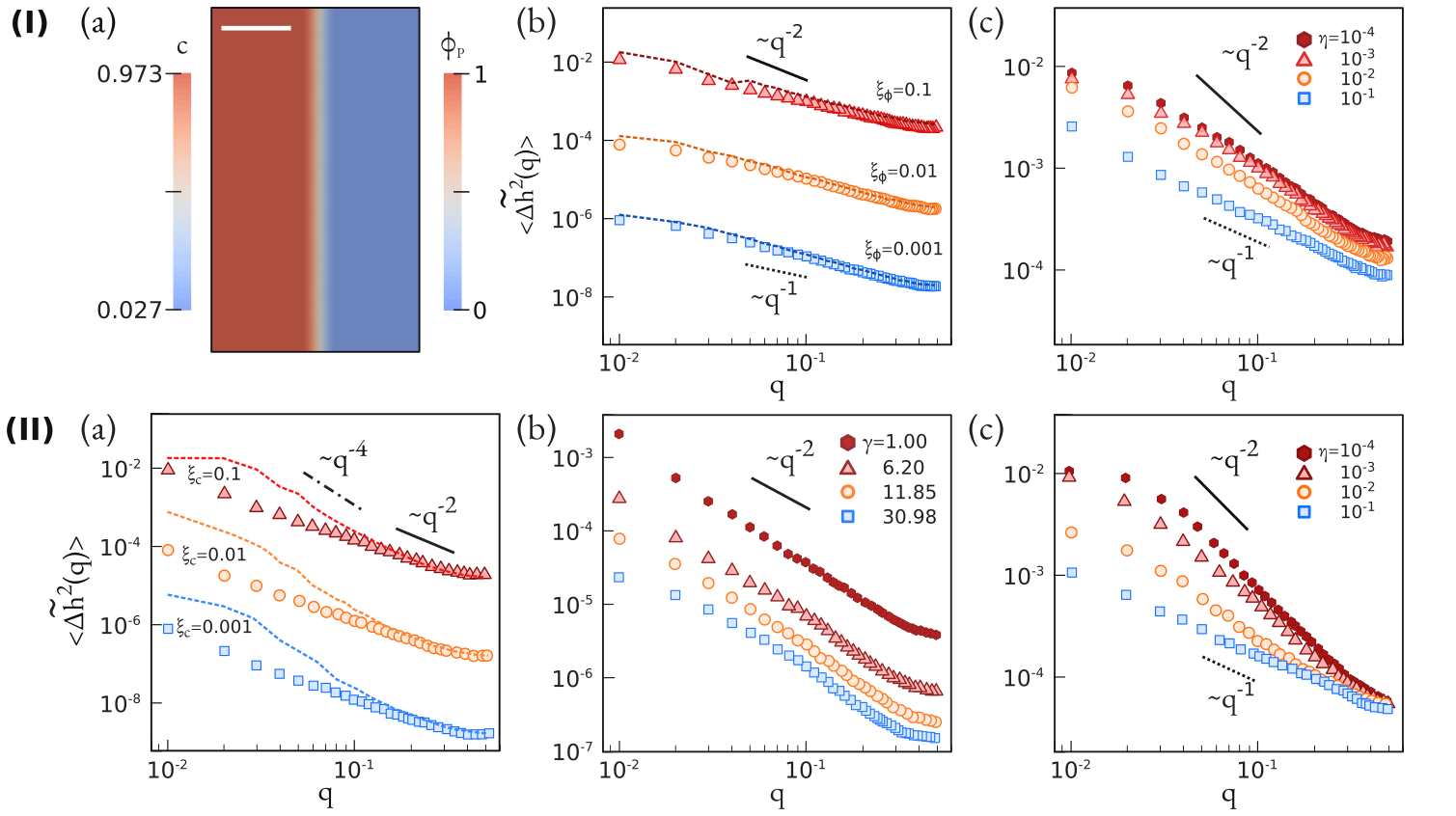}
 \centering
 \caption{Validation of the two stochastic phase-field models with the capillary wave theory. Open colored symbols: with hydrodynamics; dashed color lines: without hydrodynamics. 
 (I) Allen-Cahn-Navier-Stokes model:
  (a) The initial unperturbed flat interface.
  (b) The capillary wave amplitudes $\langle  \Delta\widetilde{h}^{\scriptscriptstyle 2}(q)\rangle$
  according to Eq.~\eqref{eq:cwt_q} in the reciprocal space for the perturbed interface with different composition noise amplitudes $\xi_{\scriptscriptstyle \phi}$. 
  (c) $\langle  \Delta\widetilde{h}^{\scriptscriptstyle 2}(q)\rangle$ with distinct viscosities $\eta$,
  for a fixed value of noise amplitude, $\xi_{\scriptscriptstyle \phi}=0.1$. 
 (II) Cahn-Hilliard-Navier-Stokes model: 
  (a) The capillary wave amplitudes increasing with $\boldsymbol\xi_{\scriptscriptstyle c}$.
   (b) $\langle  \Delta\widetilde{h}^{\scriptscriptstyle 2}(q)\rangle$ versus $q$ for
 different interfacial tensions, $\gamma$ at noise amplitude $\boldsymbol\xi_{\scriptscriptstyle c}=0.05$.  
  (c) $\langle  \Delta\widetilde{h}^{\scriptscriptstyle 2}(q)\rangle$ versus $q$ for different viscosities $\eta$ with noise amplitude $\boldsymbol\xi_{\scriptscriptstyle c}=0.1$.
 The black solid lines guide the capillary wave theory relationship of $\langle  \Delta\widetilde{h}^{\scriptscriptstyle 2}(q)\rangle\sim q^{\scriptscriptstyle -2}$. 
  The dotted, solid, and dot-dashed lines show different scaling laws, $\langle  \Delta\widetilde{h}^{\scriptscriptstyle 2}(q)\rangle\sim q^{\scriptscriptstyle -1}$, $\sim q^{\scriptscriptstyle -2}$, and $\sim q^{\scriptscriptstyle -4}$, respectively.
 }
 \label{fig:valid_cwt}
\end{figure*}

To validate the energy dissipation behaviors 
of the ACNS and CHNS models with the capillary wave theory (CWT), 
an initial setup with a flat liquid-matrix interface 
is demonstrated in Fig.~\ref{fig:valid_cwt}(I)(a). 
The density $\rho=0.01$,
the interfacial tension $\gamma=1.0$ are adopted for both models.
Firstly, for the ACNS model, 
two scenarios, namely, i) the Allen-Cahn model with hydrodynamics (open dots) 
and ii) without hydrodynamics (dashed lines) are considered 
and shown in Fig.~\ref{fig:valid_cwt}. 
The phase-field kinetic parameter $\tau$ in the ACNS model is set to be $0.01$ and the viscosity is $\eta=0.001$. 
Fig.~\ref{fig:valid_cwt}(I)(b) shows the squared capillary wave amplitude $ \big\langle \Delta\widetilde{h}^{\scriptscriptstyle 2}(q)\big\rangle$ with the wave number $q$ for different composition noise amplitudes $\xi_{\scriptscriptstyle\phi}$.
The scaling law with $q^{\scriptscriptstyle -1}<\big\langle \Delta\widetilde{h}^{\scriptscriptstyle 2}(q)\big\rangle< q^{\scriptscriptstyle -2}$ appears in both scenarios irrespective of hydrodynamics, 
as guided by the black lines. 
With the growth of the noise amplitude $\xi_{\scriptscriptstyle \phi}$ by $10$ times,  $\big\langle \Delta\widetilde{h}^{\scriptscriptstyle 2}(q)\big\rangle$
increases accordingly by $10^{\scriptscriptstyle 2}$ times, 
showing good consistence with Eq.~\eqref{eq:cwt_q}. 
Fig.~\ref{fig:valid_cwt}(I)(c) illustrates the impact of the viscosity $\eta=\eta_{\scriptscriptstyle P}=\eta_{\scriptscriptstyle M}$ on 
$\big\langle \Delta\widetilde{h}^{\scriptscriptstyle 2}(q)\big\rangle$ at a fixed noise amplitude $\xi_{\scriptscriptstyle \phi}=0.1$.
Not only the capillary wave amplitude decreases with larger viscosity, 
but the scaling factor reduces to $-1$ at large viscosity.
This scaling law does not obey CWT,
implying that the stochastic ACNS model 
embodies a distinct energy dissipation mechanism 
for the perturbed interface 
from the binary fluid systems.

Next, the CHNS model is validated and the results are shown 
in Fig.~\ref{fig:valid_cwt}(II). 
Here, 
the structure factor $\big\langle \Delta\widetilde{h}^{\scriptscriptstyle 2}(q)\big\rangle$ is also measured for two scenarios: 
i) with hydrodynamics (open dots) 
and ii) without hydrodynamics (dashed lines). 
The mobility $\mathcal{M}_0$ set to be $1.0$ 
and the viscosity is $\eta=0.001$. 
In good consistency with previous researches~\cite{davis1977capillary}, 
at large wave number $q$, 
$\big\langle \Delta\widetilde{h}^{\scriptscriptstyle 2}(q)\big\rangle$ 
shows the $q^{\scriptscriptstyle -2}$ tendency with the wave number $q$ for both scenarios, 
as guided by the black solid line in Fig.~\ref{fig:valid_cwt}(II)(a). 
For the short wavelength perturbations with $q<0.1$, 
the Cahn-Hilliard model without hydrodynamics (dashed colored lines)
shows an apparent deviation from the CWT scaling law. 
It indicates that hydrodynamics is a crucial mechanism for the surface energy dissipation of the fluid interface, 
especially for the noise with short wavelengths. 
Then, in Fig.~\ref{fig:valid_cwt}(II)(b), 
we fix the constant noise amplitude $\xi_{\scriptscriptstyle c}=0.05$,
and observe the reduction in capillary wave amplitudes with the increase in the surface tension $\gamma$ 
which is in good accordance with the CWT.
In Fig.~\ref{fig:valid_cwt}(II)(c), 
the viscosity effect on 
the scaling law of $\big\langle \Delta\widetilde{h}^{\scriptscriptstyle 2}(q)\big\rangle$ versus $q$ is illustrated.
By setting a larger $\eta$, 
the viscous stress $\eta\nabla^{\scriptscriptstyle 2}\boldsymbol{u}$ in the Navier-Stokes equation 
is magnified, giving rise to a stronger energy dissipation via frictional forces between fluids.
But the deduction of Eq.~\eqref{eq:cwt_q} only considers the surface energy dissipation 
and does not take the viscosity effect on the kinetic energy into account.
This observation in turn 
indicates that for the composition noise dominated Brownian motion of sub-micro droplets, 
the viscous effect is of subtle importance.
In this way, we set low viscosity $\eta=0.001$ in the following parts to eliminate the viscous dissipation mechanism, 
which is in line with the CWT and experiments.

\subsection{Dispersion relation}
To have a better understanding of the capillary wave theory for two types of phase-field models, 
we scrutinize the energy dissipation by the dispersion relation which explains the distinct scaling laws between the capillary wave amplitude $\langle\Delta^{\scriptscriptstyle 2}h\rangle$ with the wavenumber $q$.
Here, we suggest two regimes with distinct dispersion relations, namely,  composition noise dominated regime and convection dominated regime. 
\subsubsection{\textcolor{black}{Composition noise dominated regime}}
(I) For the ACNS model, the thermal fluctuation energy-gaining rate of the system reads
\begin{align*}
    \frac{\partial \mathcal{F}^{\scriptscriptstyle +}}{\partial t} &= \int_{\scriptscriptstyle \Omega} f^{\scriptscriptstyle *}\tau\,\xi_{\scriptscriptstyle \phi}^{\scriptscriptstyle 2}\, d\Omega
    =f^{\scriptscriptstyle *}\tau_{\scriptscriptstyle 0}\,\xi_{\scriptscriptstyle \phi}^{\scriptscriptstyle 2}\,\text{S},
\end{align*}
where $\text{S}$ is the interface area and the characteristic energy density $f^{\scriptscriptstyle *}=\gamma/\epsilon$. 
The energy dissipation rate by the diffusion process obeys the following energy law as

\begin{align*}
    \frac{\partial \mathcal{F}^{\scriptscriptstyle -}}{\partial t}&= \int_{\scriptscriptstyle \Omega} \frac{\delta F}{\delta\phi}\frac{d\phi}{dt} d\Omega
    = -\!\!\int_{\scriptscriptstyle \Omega} \tau\,\mu^{\scriptscriptstyle 2} d\Omega\\
    &=-\!\!\int_{\scriptscriptstyle \Omega} \tau\Big[\gamma(1-2\phi)-\gamma\nabla^{\scriptscriptstyle 2}\phi\Big]^{\scriptscriptstyle 2} d\Omega\\
    &=\text{S}\int_{\scriptscriptstyle -\infty}^{\scriptscriptstyle \infty}\!\! \bar \alpha\,\tau\,f^{\scriptscriptstyle * 2}\,\big(1+2\pi^{\scriptscriptstyle 2}\epsilon^{\scriptscriptstyle 2}\,q^{\scriptscriptstyle 2}+\pi^{\scriptscriptstyle 4}\epsilon^{\scriptscriptstyle 4}q^{\scriptscriptstyle 4}\big) \bar\phi^{\scriptscriptstyle\,  2}dx \\
    &=\bar\alpha\,\tau_{\scriptscriptstyle 0}\,f^{\scriptscriptstyle * 2}\,\text{S}\,\big(1+2\pi^{\scriptscriptstyle 2}\epsilon^{\scriptscriptstyle 2}\,q^{\scriptscriptstyle 2}+\pi^{\scriptscriptstyle 4}\epsilon^{\scriptscriptstyle 4}q^{\scriptscriptstyle 4}\big)\Delta^{\scriptscriptstyle 2} h.
\end{align*}
Here, we expand the composition at the interface location $\phi=0.5$ to the first order of infinitesimal length $\bar\epsilon$ 
and integrate by substituting $\phi=0.5+\bar\epsilon\,\bar\phi$.  
With tiny thermal noises, the composition perturbation $\bar\phi$ is described by a wave function $e^{\scriptscriptstyle \zeta t-iqy}$. 
Therefore, the terms $\nabla\bar\phi$ and $\nabla^{\scriptscriptstyle 2}\bar\phi$ are linearized as $-i\,q\,\bar\phi$ and $q^{\scriptscriptstyle 2}\,\bar\phi$, respectively. 
In addition, we rewrite $\bar\phi=\Delta\phi=\nabla\phi\cdot \Delta h$ and let $\bar \alpha= 16\,\bar\epsilon^{\scriptscriptstyle \,2}/\pi^{\scriptscriptstyle 4}$. 
Moreover, the first-order approximation simplifies $\int_{\scriptscriptstyle -\infty}^{\scriptscriptstyle \infty}(\nabla \phi)^{\scriptscriptstyle 2}dx$ by the Dirac's delta function. 
In this way, at equilibrium, we have $\partial_{\scriptscriptstyle t} F^{\scriptscriptstyle +}=\partial_{\scriptscriptstyle t} F^{\scriptscriptstyle -}$ which denotes the balance between energy gaining and consumption, so that
\begin{align}
    \!\!\!\Delta^{\scriptscriptstyle 2} h \!\propto \!\frac{\xi_{\scriptscriptstyle \phi}^{\scriptscriptstyle 2}}{\gamma\,\big(1+2\pi^{\scriptscriptstyle 2}\epsilon^{\scriptscriptstyle 2}\,q^{\scriptscriptstyle 2}+\pi^{\scriptscriptstyle 4}\epsilon^{\scriptscriptstyle 4}q^{\scriptscriptstyle 4}\big)}
    \label{eq:dispersion_ac}.
\end{align}
With this dispersion relation, we address the scaling law of the capillary wave amplitudes
observed in Fig.~\ref{fig:valid_cwt}(I). 
Distinct from the Fick's second law and the Cahn-Hilliard equation, 
the Allen-Cahn equation is a diffusion-reaction equation and can be expressed with the composition perturbation $\bar \phi$ as. 
\begin{align*}
    \frac{d \bar \phi}{dt}=\tau\Big(\frac{16\gamma}{\pi^{\scriptscriptstyle 2}\epsilon}\bar \phi - \gamma\epsilon\,\nabla^{\scriptscriptstyle 2} \bar \phi\Big).
\end{align*}
For small wavelength noises, the linear term of $\bar \phi$ controls the energy dissipation over the curvature related $\nabla^{\scriptscriptstyle 2}\bar\phi$ term. 
While for the long wavelength perturbations, 
the second-order term comes into play, 
resulting the scaling factor of $\langle\Delta^{\scriptscriptstyle 2}h\rangle$ approaching $-2$.

(II) For the CHNS model, the fluctuation energy-gaining rate reads
\begin{align*}
    \frac{\partial \mathcal{F}^{\scriptscriptstyle +}}{\partial t} = \int_{\scriptscriptstyle \Omega} f^{\scriptscriptstyle *}\mathcal{M}\,\xi_{\scriptscriptstyle c}^{\scriptscriptstyle 2}\, d\Omega
    = f^{\scriptscriptstyle *}\mathcal{M}_{\scriptscriptstyle 0}\,\xi_{\scriptscriptstyle c}^{\scriptscriptstyle 2}\,\text{S}.
\end{align*}
Meanwhile, the energy dissipates as
\begin{align*}
    \frac{\partial \mathcal{F}^{\scriptscriptstyle -}}{\partial t}&= \int_{\scriptscriptstyle \Omega} \frac{\delta F}{\delta c}\frac{dc}{dt} d\Omega\\
    &= -\!\!\int_{\scriptscriptstyle \Omega} \mathcal{M}(\nabla\mu)^{\scriptscriptstyle 2} d\Omega\\
    &=\text{S}\int_{\scriptscriptstyle -\infty}^{\scriptscriptstyle \infty}\!\! \mathcal{M}\,f^{\scriptscriptstyle * 2}\,\bar\epsilon^{\scriptscriptstyle \,2}\,\big(q^{\scriptscriptstyle 2}\chi^{\scriptscriptstyle 2}+2\epsilon^{\scriptscriptstyle 2}q^{\scriptscriptstyle 4}\chi+ \epsilon^{\scriptscriptstyle 4}q^{\scriptscriptstyle 6}\big)\,\bar c ^{\scriptscriptstyle \,2}\,dx\\
    &=\mathcal{M}_{\scriptscriptstyle 0}\,f^{\scriptscriptstyle * 2}\,\bar\epsilon^{\scriptscriptstyle \,2}\,\text{S}\,\big(q^{\scriptscriptstyle 2}\chi^{\scriptscriptstyle 2}+2\epsilon^{\scriptscriptstyle 2}q^{\scriptscriptstyle 4}\chi+ \epsilon^{\scriptscriptstyle 4}q^{\scriptscriptstyle 6}\big)\Delta^{\scriptscriptstyle 2} h,
\end{align*}
in which the integrated term is expanded at the interface position with $c = 0.5 + \bar \epsilon\,\bar c$. 
Similar to the mathematical treatments in the ACNS model, 
the composition perturbation $\bar c$ is analogized with the wave function $e^{\scriptscriptstyle \zeta t-iqy}$.
At equilibrium, we obtain
\begin{align}
    \Delta^{\scriptscriptstyle 2} h \propto \frac{\xi_{\scriptscriptstyle c}^{\scriptscriptstyle 2}}{\gamma\,\big(\chi^{\scriptscriptstyle 2}q^{\scriptscriptstyle 2}+2\chi \,\epsilon^{\scriptscriptstyle 2}q^{\scriptscriptstyle 4}+ \epsilon^{\scriptscriptstyle 4}q^{\scriptscriptstyle 6}\big)}.
\end{align}
Both $q^{\scriptscriptstyle -2}$ and $q^{\scriptscriptstyle -4}$ scaling laws are captured in the simulated CWT of the Cahn-Hilliard model without hydrodynamics; see dashed lines in Fig.~\ref{fig:valid_cwt}(II)(a). 
It reflects the prominent difference between the Cahn-Hilliard equation and the diffusion equation (Fick's second law), 
since CH is a fourth-order partial differential equation expressed with composition perturbation $\bar c$ as
\begin{align*}
    \frac{d \bar c}{dt}=\nabla\cdot(\mathcal{M}\frac{\sigma\chi}{\epsilon}\nabla \bar c) - \nabla\cdot(\mathcal{M}\sigma\epsilon\,\nabla^{\scriptscriptstyle 3} \bar c).
\end{align*}
Hence, the energy dissipation for small wavelength noises (large $q$) behaves similarly to the standard diffusion process. 
While for the large wavelength (small $q$), 
its dissipation is dominated by
the fourth-order term $\sigma\nabla^{\scriptscriptstyle 4}\bar c$. 

\subsubsection{Convection dominated regime}
The energy law behaves entirely differently 
when convection overwhelms diffusion, 
and we have another energy law for the ACNS model
\begin{align*}
    \frac{\partial \mathcal{F}^{\scriptscriptstyle -}}{\partial t} &= \int_{\scriptscriptstyle \Omega} \boldsymbol{u}\cdot\rho\frac{d \boldsymbol{u}}{d t} \,d\Omega
    = \!\!\int_{\scriptscriptstyle \Omega} \boldsymbol{u} \cdot\big(\mu\nabla c\big)\, d\Omega\\
    &=\text{S}\int_{\scriptscriptstyle -\infty}^{\scriptscriptstyle \infty}\!\!\nabla\Psi\cdot\frac{16 f^{\scriptscriptstyle *}}{\pi^{\scriptscriptstyle 2}}\big(-\bar\epsilon\,\bar \phi-\pi^{\scriptscriptstyle 2}\epsilon^{\scriptscriptstyle 2}\,\bar\epsilon\,\nabla^{\scriptscriptstyle 2}\bar \phi\big)\big( \bar\epsilon\,\nabla \bar \phi\big) dx \\
    &=\frac{16 f^{\scriptscriptstyle *}\, \text{S}}{\pi^{\scriptscriptstyle 2}}\bar\epsilon^{\scriptscriptstyle \,2}q^{\scriptscriptstyle \prime}\Psi\,\big(\chi\,q + \pi^{\scriptscriptstyle 2}\epsilon^{\scriptscriptstyle 2}q^{\scriptscriptstyle 3} \big)\Delta^{\scriptscriptstyle 2} h.
\end{align*}
Here, we assume the velocity $\boldsymbol{u}$ as the gradient of the tiny perturbed stream function $\Psi=e^{\scriptscriptstyle \zeta^{\prime}t-iq^{\prime}x}$ with the different phase parameter $q^{\scriptscriptstyle \prime}$ from the composition noise.
Under this circumstance, the capillary wave amplitude is derived as follows
\begin{align}
    \Delta^{\scriptscriptstyle 2} h \propto \frac{\tau_{\scriptscriptstyle 0}\,\xi_{\scriptscriptstyle \phi}^{\scriptscriptstyle 2}}{ q+\pi^{\scriptscriptstyle 2}\,\epsilon^{\scriptscriptstyle 2} \,q^{\scriptscriptstyle 3}}.
\end{align}
Comparing with Eq.~\eqref{eq:dispersion_ac}, $\langle\Delta^{\scriptscriptstyle 2} h\rangle$ has a scaling factor between $-1$ and $-2$ with respect to the wavenumber $q$ in the ACNS model. 
Similarly, the energy law for the convection-dominated CHNS model is deduced as 
\begin{align}
    \Delta^{\scriptscriptstyle 2} h \propto \frac{\mathcal{M}_{\scriptscriptstyle 0}\,\xi_{\scriptscriptstyle c}^{\scriptscriptstyle 2}}{\chi\, q+\epsilon^{\scriptscriptstyle 2} \,q^{\scriptscriptstyle 3}},\label{eq:dispersion_hd_ch}
\end{align}
which is also in line with the CWT simulation results shown in Fig.~\ref{fig:valid_cwt}(II). 
These dispersion relations are dealing with the composition noise dissipated via convection, 
and has never been considered in previous FDT and CWT theories.
We stress that this energy dissipation mechanism is entirely different from the one in the Langevin mechanics, where the random body force gets smoothed by the viscous stress.
Testified in previous simulations~\cite{shang2011fluctuating,chaudhri2014modeling}, 
the random body force perturbed interface still follows the CWT scaling law with $\langle\Delta^{\scriptscriptstyle 2} h\rangle\sim q^{\scriptscriptstyle -2}$.
\section{Result and discussion}
In this section, we present the Brownian motion simulations 
with both models of CHNS and ACNS.
Different noise amplitudes, droplet radius, and 
particle-matrix interfacial tension are considered,
and their impacts on Brownian behaviors are discussed. 
\subsection{Equilibrium behaviors}\label{sec:equi}
\subsubsection{The Einstein's relation}
\begin{figure*}[htbp]
 \includegraphics[width=0.86\linewidth]{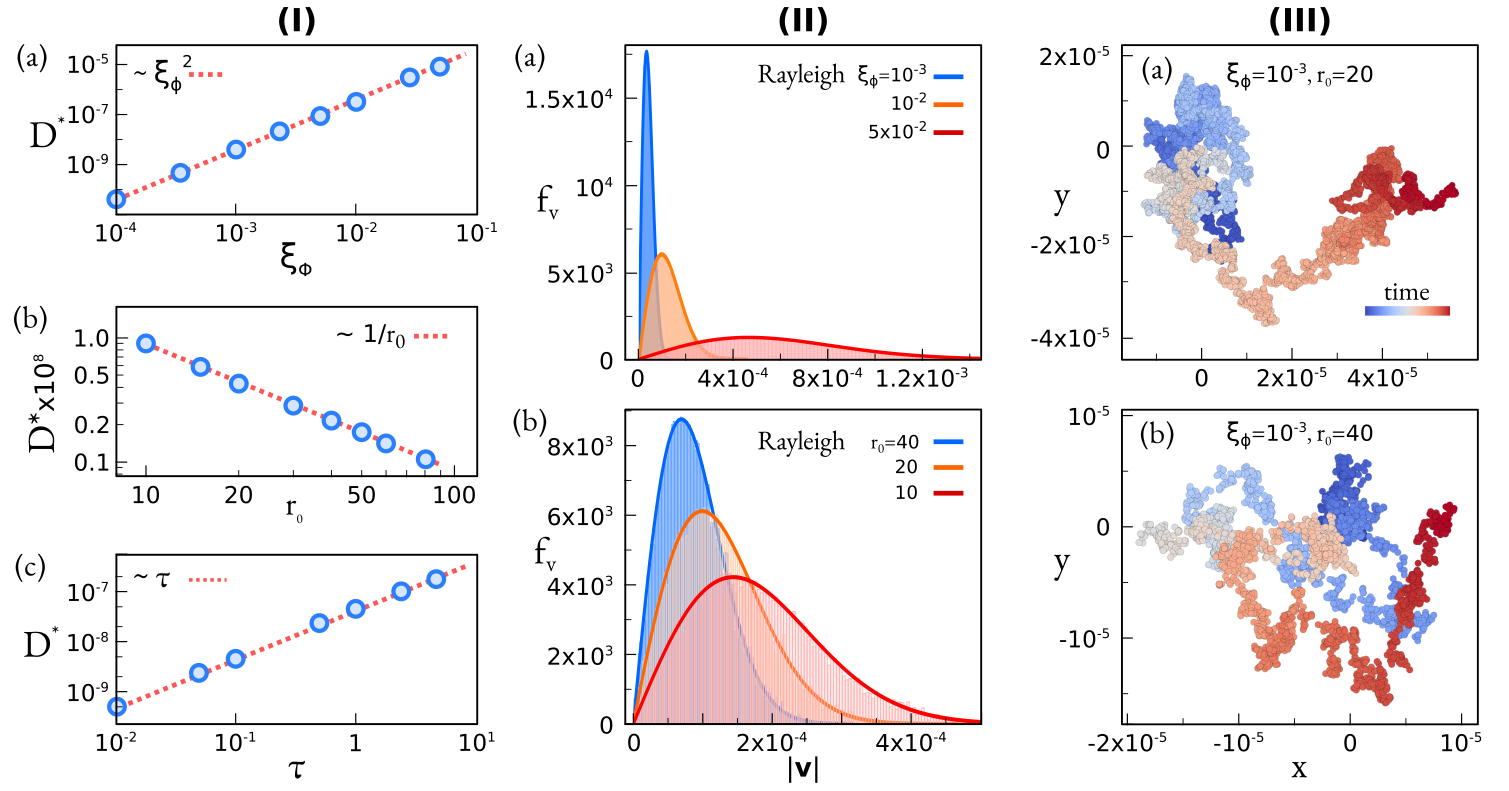}
 \centering
 \caption{2D Brownian motion of droplets via ACNS model. 
 (I) The Brownian motion coefficient $D^*$,
 (a) with noise amplitude $\boldsymbol\xi_{\scriptscriptstyle \phi}$;
 (b) with initial radius $r_{\scriptscriptstyle 0}$;
 (c) with kinetic parameter $\tau$.
 The dashed red lines guide the $D^* \propto\xi_{\scriptscriptstyle \phi}^{\scriptscriptstyle 2}$, $1/r_{\scriptscriptstyle0}$, 
 and $1/\tau$ relations.
 (II) Probability density function (PDF) of particle velocity follows the Rayleigh distribution,
 (a) with $\boldsymbol\xi_{\scriptscriptstyle \phi}$;
 (b) with $r_{\scriptscriptstyle 0}$. 
 (III) The particle trajectory.
 The color bar scales the time sequence.
 }
 \label{fig:einstein_ac}
\end{figure*}
We present a simple proof of the Einstein relation replicated with the phase-field model. 
The total fluctuation energy normalized by the characteristic chemical free energy density $f^{\scriptscriptstyle *}$ on the 2-dimensional particle with radius $r_{\scriptscriptstyle 0}$ reads
\begin{align*}
   \langle E \rangle / f^*
   &=\int_{\scriptscriptstyle \Omega} \langle\sqrt{\tau}\,\xi_{\scriptscriptstyle \phi}, \sqrt{\tau}\,\xi_{\scriptscriptstyle \phi}^{\scriptscriptstyle \prime}\rangle \, d\Omega\\
   &=\,\tau_{\scriptscriptstyle 0}\,\xi_{\scriptscriptstyle \phi}^{\scriptscriptstyle 2}\int_{\scriptscriptstyle 0}^{\scriptscriptstyle \infty}\!\! \phi\,(1-\phi)\,2\pi r \,dr\\
   &= \, \tau_{\scriptscriptstyle 0}\,\xi_{\scriptscriptstyle \phi}^{\scriptscriptstyle 2}\,\text{S},
\end{align*}
where $\text{S}=\int_{\scriptscriptstyle 0}^{\scriptscriptstyle\infty}\phi\,(1-\phi)\,2\pi\, r dr=2\pi r_{\scriptscriptstyle 0}$ represents the surface area of the perfect spherical droplet with the interface width approaching the sharp interface limit~\cite{choudhury2012grand}. 
The characteristic chemical free energy $f^*$ equates to $1.0$ after non-dimensionalization.
Consequently, with the property of the Rayleigh distribution, the root mean square droplet velocity in 2D reads
\begin{align*}
    v_{\scriptscriptstyle rms}=\sqrt{\langle \boldsymbol v^{\scriptscriptstyle 2}\rangle}
    = \sqrt{\frac{2\langle E \rangle}{m}} 
    =\sqrt{\frac{4\, \tau_{\scriptscriptstyle 0}\,\xi_{\scriptscriptstyle \phi}^{\scriptscriptstyle 2}}{r_{\scriptscriptstyle 0}}}
    =\sqrt{2 D^*_{\scriptscriptstyle AC}},
\end{align*}
from which the Brownian coefficient $D^*$ of the ACNS model follows
\begin{align}
    D^*_{\scriptscriptstyle AC}=
    \frac{2\,\tau_{\scriptscriptstyle 0} \,\xi_{\scriptscriptstyle \phi}^{\scriptscriptstyle 2}}{r_{\scriptscriptstyle 0}}.\label{eq:d_ac}
\end{align}
With the same method, the Brownian coefficient with the CHNS model reads
\begin{align}
    D^*_{\scriptscriptstyle CH}= 
    \frac{2\,\mathcal{M}_{\scriptscriptstyle 0}\,\xi_{\scriptscriptstyle c}^{\scriptscriptstyle 2}}{r_{\scriptscriptstyle 0}}.\label{eq:d_ch}
\end{align}

To testify the Einstein's relation, 
we perform 2D Brownian motion simulations 
with the ACNS model first. 
A droplet with an initial radius of
$r_{\scriptscriptstyle 0}=20$ is placed amid the domain with a size of $12r_{\scriptscriptstyle 0}\times 12r_{\scriptscriptstyle 0}$
which can eliminate the influence of boundary 
on the droplet motion~\cite{mo2019highly} (see supplementary). 
With an increase in the noise amplitude $\xi_{\scriptscriptstyle \phi}$, 
the Brownian coefficient $D^*$ shows 
a parabolic relation with $\xi_{\scriptscriptstyle \phi}$, 
as guided by the red dashed line in Fig.~\ref{fig:einstein_ac}(I)(a). 
According to Eq.~\eqref{eq:noise_phi}, the linear dependency of $D^*$ on $k_{\scriptscriptstyle B} T$ is confirmed.
Here, $D^*$ is fitted with the Rayleigh distribution based on the droplet velocity distributions for $10^{\scriptscriptstyle 6}$ time steps, 
as sketched in Fig.~\ref{fig:einstein_ac}(II)(a). 
For stronger noise (or higher temperatures), 
the probability distribution function (PDF) of velocity becomes broad 
and shifts to the high-velocity side, 
indicating enhanced droplet motion by the composition noise.

Next, we alter the droplet radius $r_{\scriptscriptstyle 0}$ for 
a constant noise amplitude $\xi_{\scriptscriptstyle \phi}=0.001$.
The inverse relationship of $D^* \sim 1/r_{\scriptscriptstyle 0}$ is clearly demonstrated in the mid row of Fig.~\ref{fig:einstein_ac}(I)(b), as guided by the red dashed line. 
By increasing the droplet radius, 
we observe that the peak of the velocity PDF moves to the low-velocity in Fig.~\ref{fig:einstein_ac}(II)(b).
It implies that 
the Brownian droplet approaching its equilibrium 
is influenced by the size effect. 
Moreover, we observe another linear relationship between $D^*$ and the kinetic parameter $\tau$ 
which is shown in the lower row of Fig.~\ref{fig:einstein_ac}(I)(c). 
Compared to the 2D droplet trajectories in Fig.~\ref{fig:einstein_ac}(III), 
the larger molecular mobility not only enhances the macroscopic diffusion of the whole Brownian droplet 
but also modifies the motion behavior which is discussed later in Sec.~\ref{sec:pd}. 
\begin{figure*}[tbp]
 \includegraphics[width=0.86\linewidth]{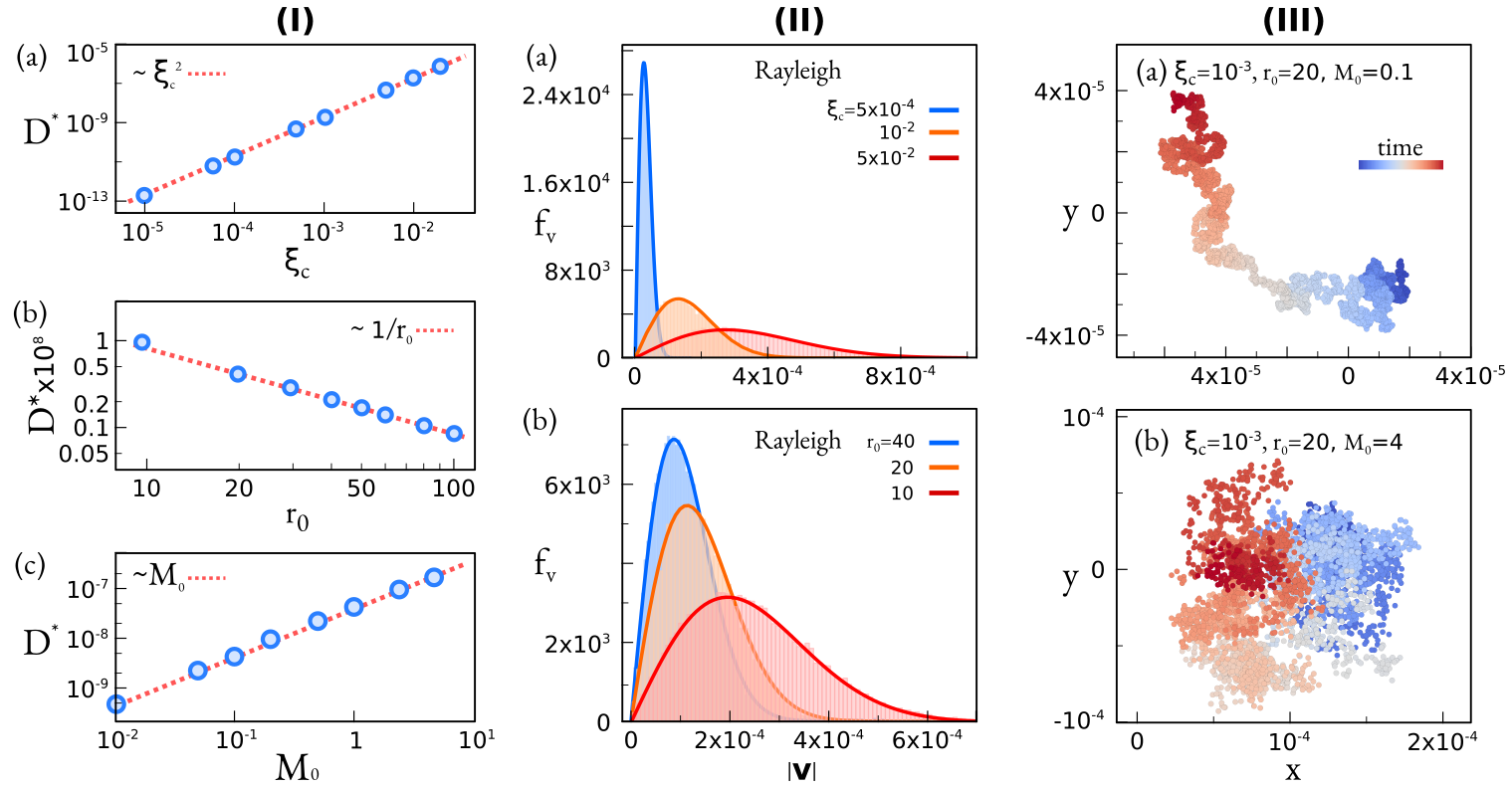}
 \centering
 \caption{2D Brownian motion of droplets with CHNS model. 
 (I) The Brownian motion coefficient $D^*$,
 (a) with noise amplitude $\boldsymbol\xi_{\scriptscriptstyle c}$;
 (b) with initial radius $r_{\scriptscriptstyle 0}$;
 (c) with mobility parameter $\mathcal{M}_{\scriptscriptstyle 0}$.
 The dashed red lines guide the $D^* \propto\xi_{\scriptscriptstyle \phi}^{\scriptscriptstyle 2}$, $1/r_{\scriptscriptstyle0}$, and $\mathcal{M}_{\scriptscriptstyle 0}$ relations.
 (II) Probability density function (PDF) of droplet velocity follows the Rayleigh distribution,
 (a) with $\boldsymbol\xi_{\scriptscriptstyle c}$;
 (b) with $r_{\scriptscriptstyle 0}$. 
 (III) The droplet trajectory.
 The color bar scales the time sequence.
 }
 \label{fig:einstein_ch}
\end{figure*}
In addition, Einstein's relation is replicated with the CHNS model. 
The initial droplet radius is $20$ 
and the surface tension parameter is set to be $\gamma=1.0$. 
The remaining hydrodynamic parameters in the Navier-Stokes equation are identical to the setup in the validation section Sec.~\ref{sec:cwt}.
The Brownian coefficient $D^*$ is obtained by fitting the droplet velocity with the Rayleigh distribution. 
Fig.~\ref{fig:einstein_ch}(I) presents
three relations with (a) $D^*\propto \xi_{\scriptscriptstyle c}^{\scriptscriptstyle 2}$, (b) $D^*\propto 1/r_{\scriptscriptstyle 0}$, and (c) $D^*\propto \mathcal{M}_{\scriptscriptstyle 0}$
for various noise amplitudes $\xi_{\scriptscriptstyle \phi}$, 
radii $r_{\scriptscriptstyle 0}$, and mobility parameter $\mathcal{M}_{\scriptscriptstyle 0}$. 
Minor differences between ACNS and CHNS are attributed to the different fluctuation-dissipation scaling laws of CWT in Sec.~\ref{sec:cwt}. 
The equilibrium composition distribution at the interface for both models can also play a role, 
as sinus and hyper-tangent functions describe the ACNS and CHNS models, respectively. 
Therefore, the kinetic parameters $\tau$ and $\mathcal{M}$ are correspondingly assigned with different values, 
which results in the distinct spatial composition noises (scaled by $\sqrt{\tau}$ or $\sqrt{\mathcal{M}}$) at the interface region.

\subsubsection{Fluctuation-dissipation theorem}\label{sec:fdt}
Taking the CHNS model as an example, 
to trigger the Brownian motion,
two thermal fluctuations are considered, namely 
the composition noise $\sqrt{\mathcal{M}}\,\boldsymbol\xi_{\scriptscriptstyle c}$ 
and the random body force $\sqrt{\eta}\,\boldsymbol{F}$. 
Accordingly, each noise has its individual dissipation mechanism, 
as $\sqrt{\mathcal{M}}\,\boldsymbol\xi_{\scriptscriptstyle c}$ smoothed by diffusion 
and $\sqrt{\eta}\,\boldsymbol{F}$ by viscous friction.
Hence, we categorize the BM into two subgroups, 
(I) the composition noise dominated BM  which appears at higher temperatures and large inter-molecular diffusivity, 
and (II) the random body force dominated BM for the rigid body. 
Each type has its individual Brownian coefficient as 
\begin{align}
    D^{\scriptscriptstyle *} 
    =\bigg \{
    \begin{array}{ll}
    k_{\scriptscriptstyle B}T\, \zeta \propto \eta^{\scriptscriptstyle -1}, & \quad\text{for rigid body}; \\[2pt]
       k_{\scriptscriptstyle B}T M\propto \mathcal{M}_{\scriptscriptstyle 0}, & \quad\text{for soft droplet}.
  \end{array}
\end{align}
For the rigid body BM, the Brownian coefficient~\cite{ladd1994numerical,nie2009fluctuating} follows the Stokes-Einstein-Sutherland relation 
where the hydrodynamic mobility $\zeta=(a\pi\eta\, r_{\scriptscriptstyle 0})^{\scriptscriptstyle -1}$ 
and the constant $a$ is decided by the geometry of the motion.
For the droplet BM in our simulation, 
the microscopic molecular mobility $\mathcal{M}$ scales the macroscopic droplet motion.

\begin{figure*}[htp]
 \includegraphics[width=0.86\linewidth]{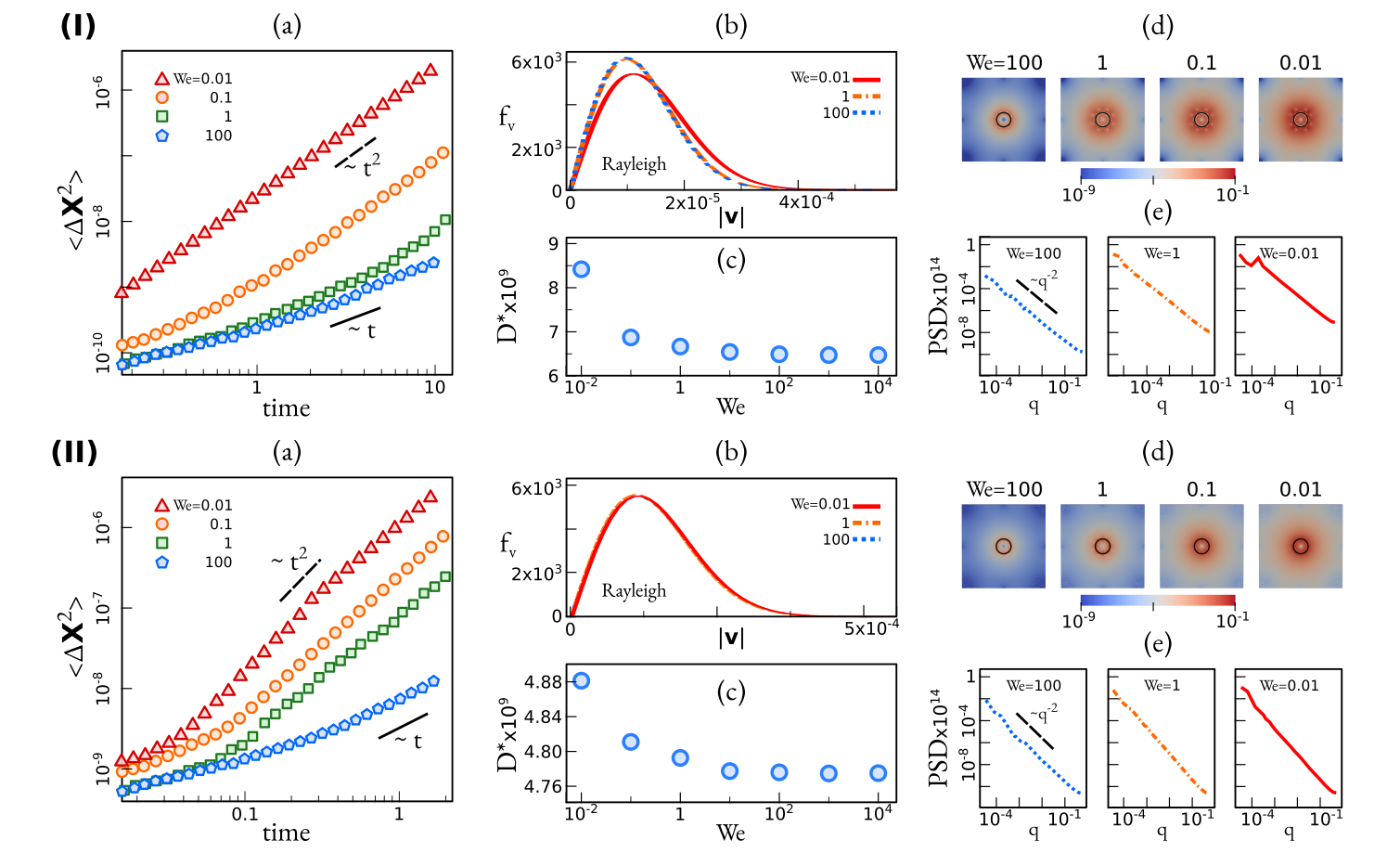}
 \centering
 \caption{2D Brownian motion of a particle 
 for increasing hydrodynamics (Weber number We) with an initial radius $r_{\scriptscriptstyle 0}=20$.
 (I) ACNS model with the composition noise amplitude $\xi_{\scriptscriptstyle \phi}= 0.001$: 
 (a) Mean squared displacement with time. 
 (b) Probability density function (PDF) of the velocity.
 (c) Brownian coefficient $D^*$ with We.
 (d) Velocity field at $t=200$. The black circles mark the particle-matrix interface with $\phi=0.5$.
  The color bar (log) scales the velocity magnitude.
 (e) Power spectral density (PSD) of the particle displacement with the wavenumber $q$ shows Brownian relationship PSD$ \propto q^{\scriptscriptstyle -2}$.
 (II) CHNS model with the composition noise amplitude $\xi_{\scriptscriptstyle c}= 0.001$. 
 }
 \label{fig:gamma}
\end{figure*}

\subsection{Non-equilibrium behaviors}\label{sec:nonequi}
\textcolor{black}{The previous section addresses the equilibrium characteristics of the CHNS and ACNS models
with which the Einstein's relation is reinstated, 
serving as a validation of our phase-field models and as an illustration of their ability to replicate thermodynamic equilibrium behavior. 
In the following section, we direct our focus to the non-equilibrium droplet behaviors 
that lie beyond the scope of thermodynamic equilibrium.}

\subsubsection{\textcolor{black}{Marangoni effect and ballistic motion}}
\label{sec:pd}
\textcolor{black}{In Einstein's derivation of the Brownian coefficient, 
the particle is idealized as a rigid body, 
and the energy dissipation occurs primarily through the viscous stresses 
exerted by the surrounding matrix.
In our model, we consider the surface tension effect of the sub-micro droplet, 
as its characteristic time scale is comparable with that of Brownian motion. 
Here, the composition-induced noise not only leads to inhomogeneous composition distributions 
at the interface region, 
but also results in the interfacial tension gradients, 
invoking the surface tension force and the Marangoni flow~\cite{schmitt2016marangoni} 
that drives the Brownian motion of the droplet. 
In our simulations, we manipulate the Weber number (We) to modulate the noise induced surface tension force 
in the Navier-Stokes equation Eq.~\eqref{eq:NS1} and Eq.~\eqref{eq:NS2} for both ACNS and CHNS models. 
As demonstrated by the velocity field snapshots in Fig.\ref{fig:gamma}(I)(c) 
and (II)(c), 
a reduction in We amplifies the surface tension force, 
enhancing the Marangoni flow around the droplets. 
Most importantly, the increasing Marangoni flow triggers a transition for the droplet motion 
from Brownian motion (associated with large We) 
to ballistic motion (associated with small We).
This transition is evident in the mean squared displacement (MSD) depicted 
in Fig.~\ref{fig:gamma}(I)(a) and (II)(a), 
where the ballistic behavior of $\langle\Delta \boldsymbol{X}^{\scriptscriptstyle 2}\rangle\sim t^{\scriptscriptstyle 2}$ is illustrated by the dashed lines. }

\textcolor{black}{Notably, we propose that the ballistic behavior in droplet dynamics arises not from the fluctuation mechanism, 
but is rather related to the dissipation mechanism. 
One piece of evidence supporting this notion can be found by the droplet velocity distribution $f(\boldsymbol v)$
in Fig.~\ref{fig:gamma}(I)(b) and (II)(b). 
We observe that an increase in Weber number exerts only minimal influence on $f(\boldsymbol v)$. 
The reduction of We from $100$ to $0.01$ only leads to 
an increase of $D^*$ by $5\%$.
However, the droplet MSD follows a fundamentally different scaling law with time. 
Another evidence is supported by the power spectral density (PSD) of the droplet displacement~\cite{huang2015effect}, 
which exhibits a $-2$ scaling law with wavenumber $q$ 
in Fig.~\ref{fig:gamma}(I)(d) and (II)(d). 
Both observations signify that the ballistic droplet motion 
stems from the same origin with the Brownian motion---composition Gaussian noise. }

\textcolor{black}{In this way, we suggest that the ballistic dynamic is originated from the dissipation mechanism 
of the noise induced surface tension force 
(which is also a kind of noise),
that is not compatible with the conventional FDT stated in Sec.~\ref{sec:fdt}.
For standard FDT description, 
the composition perturbation $\sqrt{\mathcal{M}}\,\xi_{\scriptscriptstyle c}$ emanated from the chemical free energy fluctuation is dissipated by the diffusion term $\mathcal{M}\nabla^{\scriptscriptstyle 2} \mu$, 
and the random body force $\sqrt{\eta}\,\mathbf{F}$ arising from the kinetic energy gets smoothed out by the viscous term $\eta\nabla^{\scriptscriptstyle 2} \mathbf{u}$. 
In this context, both noises are assumed to be independent with no covariance.
However, in fact, not all of the chemical free energy fluctuation is smoothed by diffusion.  
Parts of it are transformed into kinetic energy perturbation by the Kortweg-stress associated surface tension force $-c\nabla\mu$ and induces the random Marangoni flow. 
This fluid flow can either be dissipated by the viscous stress, 
or in return transforms into the composition gradient. 
Hence, a more complex dynamics comes into play 
where the dissipation of the random Marangoni flow has the characteristic time scale determined by three aspects:
(i) Péclet number, Pé decides the amount of the chemical free energy fluctuation transforming into the kinetic energy perturbations; 
(ii) Weber number, We scales the strength of Marangoni effect;
(iii) Reynolds number, Re determines the viscous dissipation magnitude. }

\textcolor{black}{Consequently, 
for large composition noise dominated scenarios with low viscous fluids, 
the viscous dissipation is incapable of smoothing the random surface tension force in its characteristic time scale,
resulting in the ballistic droplet motion.
This scenario is reminiscent of the underdamped Langevin mechanics, especially in the genre of active Brownian motion, 
where the self-propelling active particle also performs the ballistic motion 
with $\langle\Delta \boldsymbol{X}^{\scriptscriptstyle 2}\rangle\sim t^{\scriptscriptstyle 2}$~\cite{schmitt2016active}.
But the differences between our observation and ABM are apparent. 
(i) The Marangoni flow only stems from the composition fluctuation, rather than from the inhomogenous surfactant distributed around droplet~\cite{chen2009self}; 
(ii) the droplet motion is dissipated not only by the viscous friction, but also by the diffusion, 
because of the inter-conversion of chemical free energy and kinetic energy by the surface tension force.
}

\subsubsection{Phase diagram of droplet motion}

\begin{figure*}[htbp]
 \includegraphics[width=0.86\linewidth]{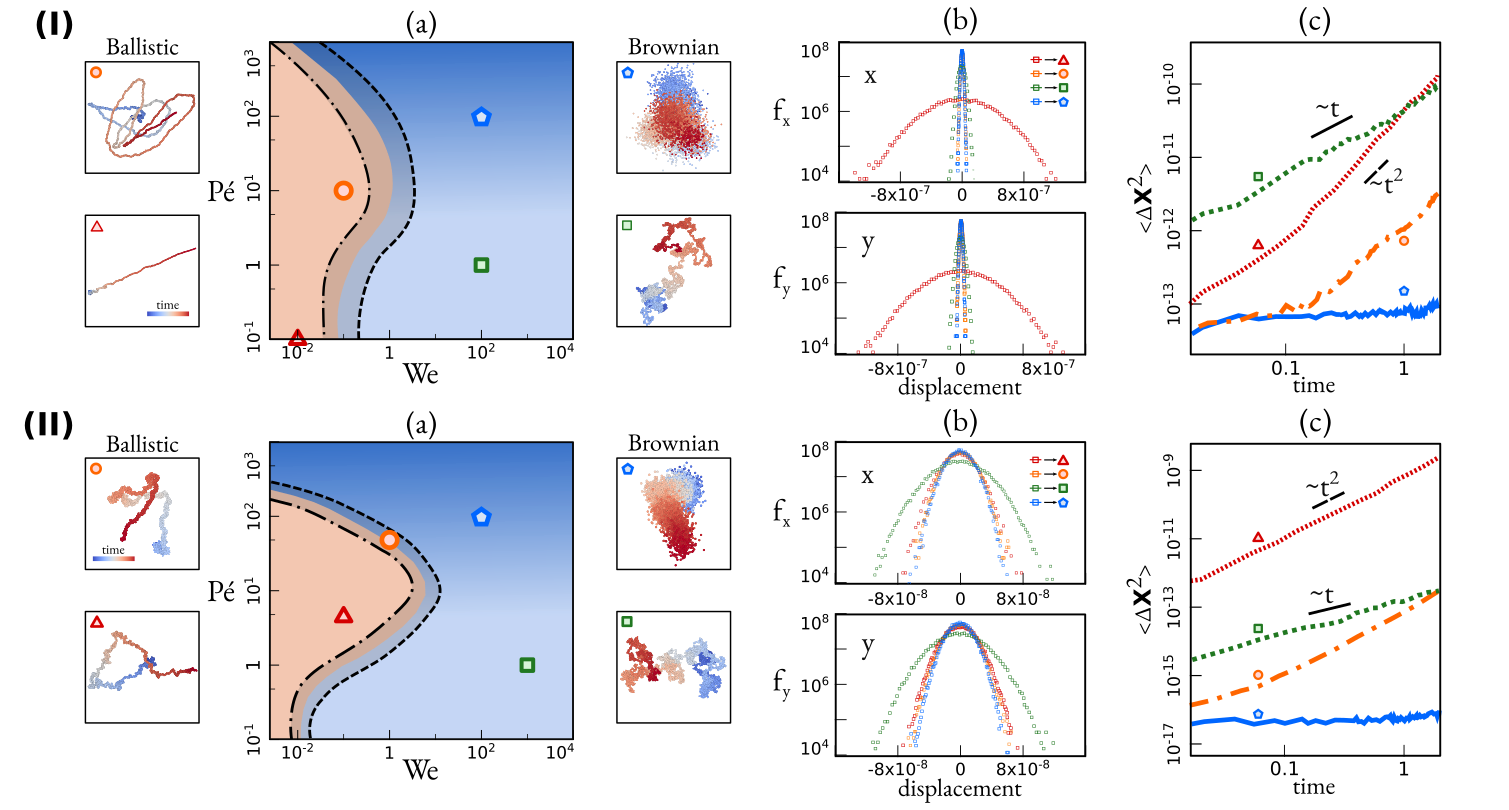}
 \centering
 \caption{The phase diagram of BM with various Weber number We and Péclet number P\text{$\acute{e}$}. 
 The initial radius is $r_{\scriptscriptstyle 0}=20$. 
 (I) ACNS model with the noise amplitude is $\xi_{\scriptscriptstyle \phi}=0.01$.
 (a) The motion phase diagram. The dot-dashed line and the dashed line separate the ballistic motion region (orange) from the no-drift Brownian region (blue) 
 and the Brownian motion with drift region (light blue).
 The light grey region in between
 the dashed and dot-dashed lines show the transition area. 
 Some exemplary motion trajectories are shown around:
 pentagon-no-drift BM;
 square-BM with drift;
 triangle-ballistic motion;
 circle-transition region.
 (b) The probability density function (PDF) of the droplet displacement. 
 (c) MSD $\langle\Delta \boldsymbol{X}^{\scriptscriptstyle 2}\rangle$ 
 of the example simulations with time.
  (II) CHNS model with the noise amplitude is $\xi_{\scriptscriptstyle c}=0.01$.
 }
 \label{fig:pd}
\end{figure*}

Upon thorough examination of droplet motion behaviors 
across various P\text{$\acute{e}$} and We values, 
we categorize droplet motion into three distinct scenarios characterized by trajectory and MSD. 

(I) Brownian motion with sub-diffusive drift: the mean squared displacement $\langle\Delta \boldsymbol{X}^{\scriptscriptstyle 2}\rangle < 2D^* t$. 
In this scenario, weak composition fluctuations are dissipated via inter-molecular diffusion,
while convection effects can be neglected. 
This scenario corresponds to the dark-blue region in the motion phase diagram of Fig.~\ref{fig:pd}(a) 
and is prominent in cases with a large P\text{$\acute{e}$}clet number and a high Weber number. 
The MSD shown by the blue pentagons in Fig.~\ref{fig:pd}(I)(b) and (II)(b) from our simulations indicates a sub-diffusive tendency in the MSD, 
deviating from the $\langle\Delta \boldsymbol{X}^{\scriptscriptstyle 2}\rangle\sim t$ behavior 
indicated by the solid line. 
Concurrently, the fourth cumulant $\langle\Delta \boldsymbol{X}^{\scriptscriptstyle 4}\rangle$ 
remains approximately at $10^{\scriptscriptstyle -32}$ without exhibiting any time dependence, 
implying that the droplet displacement adheres to a standard Gaussian distribution.

(II) Brownian motion with diffusive drift: 
convection becomes a significant factor. 
This scenario, characterized by a small P\text{$\acute{e}$} number and a large Weber number, 
is depicted by the light-blue regions in the motion phase diagram illustrated in Fig.\ref{fig:pd}. 
The composition noise, only partially dissipated through inter-molecular diffusion, 
amplifies as P\text{$\acute{e}$} decreases, 
inducing the Marangoni flow that propels the droplet's drift motion. 
Over time, this motion gradually diminishes due to the damping effect exerted by viscous stress. 
The trajectory of the droplet, observed in both models, displays a self-similar fractal structure 
and eventually reaches equilibrium, 
where the mean squared displacement (MSD) satisfies 
$\langle\Delta \boldsymbol{X}^{\scriptscriptstyle 2}\rangle\sim 2 D^* t$; see green lines in Fig.~\ref{fig:pd}(I)(c) and (II)(c). 
Concurrently, the fourth cumulant $\langle\Delta \boldsymbol{X}^{\scriptscriptstyle 4}\rangle$ 
gradually increases with time $t$. 
Over a prolonged time span, it is observed that $\langle\Delta \boldsymbol{X}^{\scriptscriptstyle 4}\rangle \sim t^{\scriptscriptstyle 2}$, akin to the findings reported in Ref.\cite{alexandre2023non}.
\begin{gather*}
    \langle\Delta \boldsymbol{X}^{\scriptscriptstyle 4} \rangle\approx 12\, \text{Var}(D^{*})\,t^{\scriptscriptstyle 2},
\end{gather*}
where $\text{Var}(D^{*})$ is the variance of  $D^*$. 
The displacement distribution observed in the simulations adheres to a Gaussian distribution, 
as depicted in Fig.~\ref{fig:pd}(b). 
This observation suggests that the time-varying non-constant Brownian coefficient $D^*$ 
is influenced by a mechanism distinct from the diffusing-diffusivity models~\cite{chubynsky2014diffusing}. 
Further comprehension of this phenomenon unveils that the deformable droplet 
deviates from its originally perfect spherical shape, 
which introduces a non-zero variance in the Brownian coefficient. 
Furthermore, since each molecule within the droplet experiences random perturbations at each time step, 
the Brownian coefficient of the entire droplet becomes a time series 
that conforms to a Gaussian distribution with Var$(D^*)>0$, 
as per the central limit theorem. 
Conversely for a rigid body, 
every molecule experiences the same noise at each time point, 
resulting in a zero variance for $D^*$.

(III) Underdamped ballistic motion:
with a further reduction in We, 
the composition fluctuation induced Marangoni flow dominates. 
Once accelerated, the droplet can be slowed down neither by the diffusion 
nor by the viscous stress,
resulting in the ballistic motion.
For both CHNS and ACNS models, 
we observe that the short-time MSD ($t<0.1$) shows the diffusion or sub-diffusion relation with time.
The droplet motion range reduces with the increase in P\text{$\acute{e}$}, 
as illustrated by the y-intercept of MSD in Fig.~\ref{fig:pd}(I)(c) and (II)(c). 
The later long-time behavior is vastly influenced by 
the composition noise induced Marangoni flow 
which is noticed by the steepening slope of MSD with a decrease in We.

\subsubsection{Fluctuation induced particle coalescence}
\textcolor{black}{In this part, 
we report another non-equilibrium behavior stemming from the thermal composition noises which give rise to a special noise induced droplet coalescence mechanism. 
Then, multi-droplet simulation is proceeded and diverse morphologies are observed due to the droplet coalescence mechanism.} 

\begin{figure*}
 \includegraphics[width=0.86\linewidth]{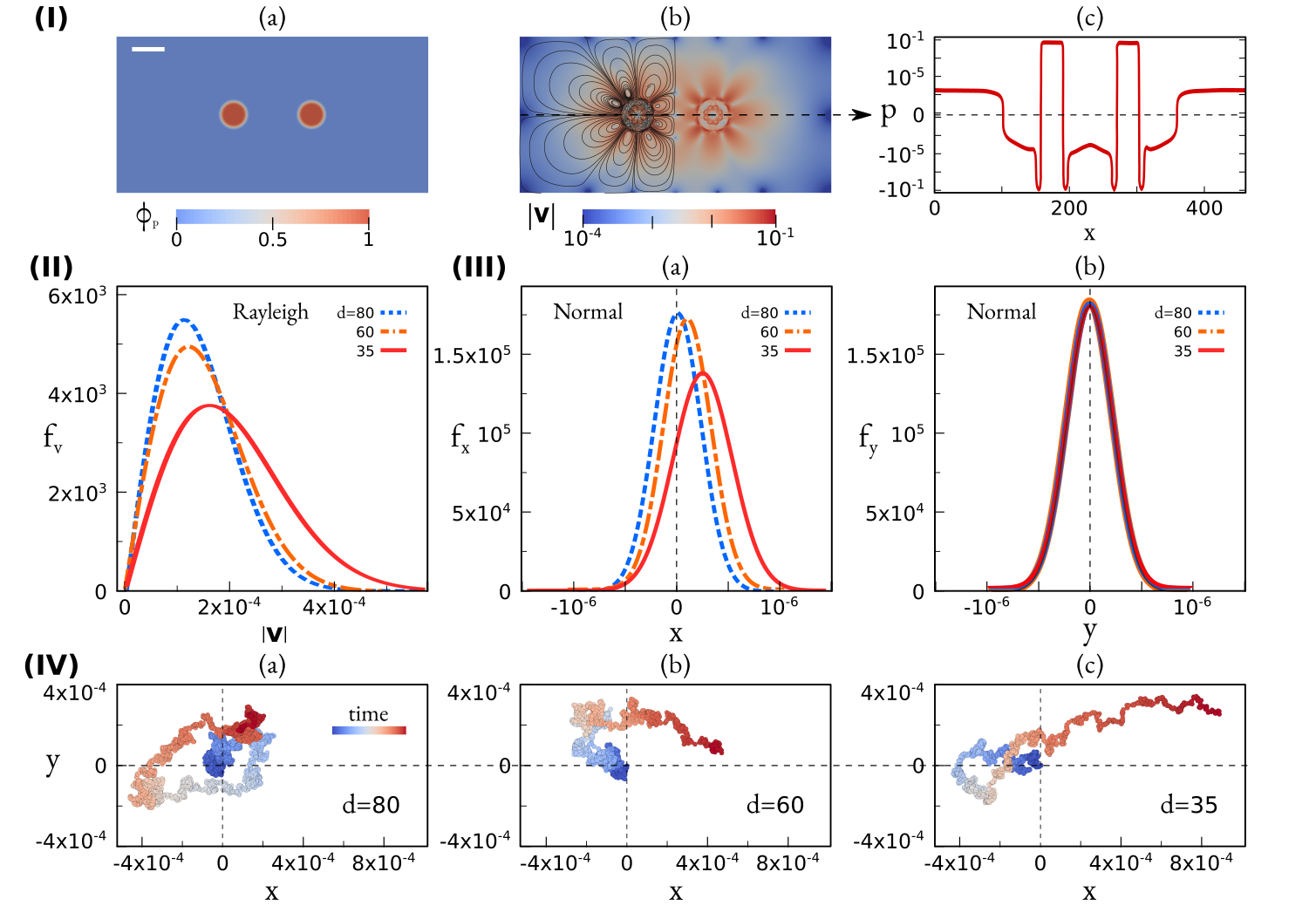}
 \centering
 \caption{2D Brownian motion of double particles with the same radius $r_{\scriptscriptstyle 0}=20$ propelled by the composition noise $\xi_{\scriptscriptstyle\phi}=0.01$.
(I) The double particle simulation with spacing at $t=200$. 
(a) $\phi_P$ particle order parameter distribution.
 The white scale bar denotes $50$.
(b) The velocity field. 
Only the left half of the streamlines are shown and the color bar (log) scales the velocity magnitude.
(c) The pressure distribution in log-scale along the line of two-particle centers .
(II) The velocity and  (III) displacement distributions of the left particle changing with $d$. 
(IV) Three trajectories of the left particle. 
The right shift manifests the deterministic motion.
The color bar scales the time sequence.
 }
 \label{fig:2p}
\end{figure*}
\textcolor{black}{Firstly, we briefly introduce the droplet coalescence mechanism
which has been extensively studied by Golovin, Tanaka~\cite{tanaka1996coarsening,golovin1995spontaneous}.
Notably, Golovin~\cite{golovin1995spontaneous} introduced 
a scenario with dual-droplets amid the surrounding matrix. An accumulation of solute material in the gap region between the droplets leads to the non-uniform solute distribution 
around the droplet,
generating a surface tension gradient and inducing Marangoni flow, 
which propels the motion of the droplets~\cite{wang2015motion}.
Golovin's diffusion-induced motion has been further investigated by Tanaka and other researchers, 
employing the Cahn-Hilliard-Navier-Stokes equation~\cite{wang2012effect}.
In the work of Tanaka~\cite{shimizu2015novel}
and also within our CHNS model,
the utilization of a double well potential energy function 
naturally results in a droplet-matrix interface 
with an infinitely wide interface region 
represented by hyperbolic tangent functions.
Hence, the concentration profiles of distinct droplets consistently overlap with one another. 
In the case of two coalescing droplets, 
this overlapping leads to an asymmetric diffusion pattern 
around each droplet, 
subsequently causing a non-uniform pressure distribution 
in the Navier-Stokes equation, giving rise to center-to-center droplet motion, 
and facilitating their coalescence.}

\begin{figure*}[htbp]
 \includegraphics[width=0.86\linewidth]{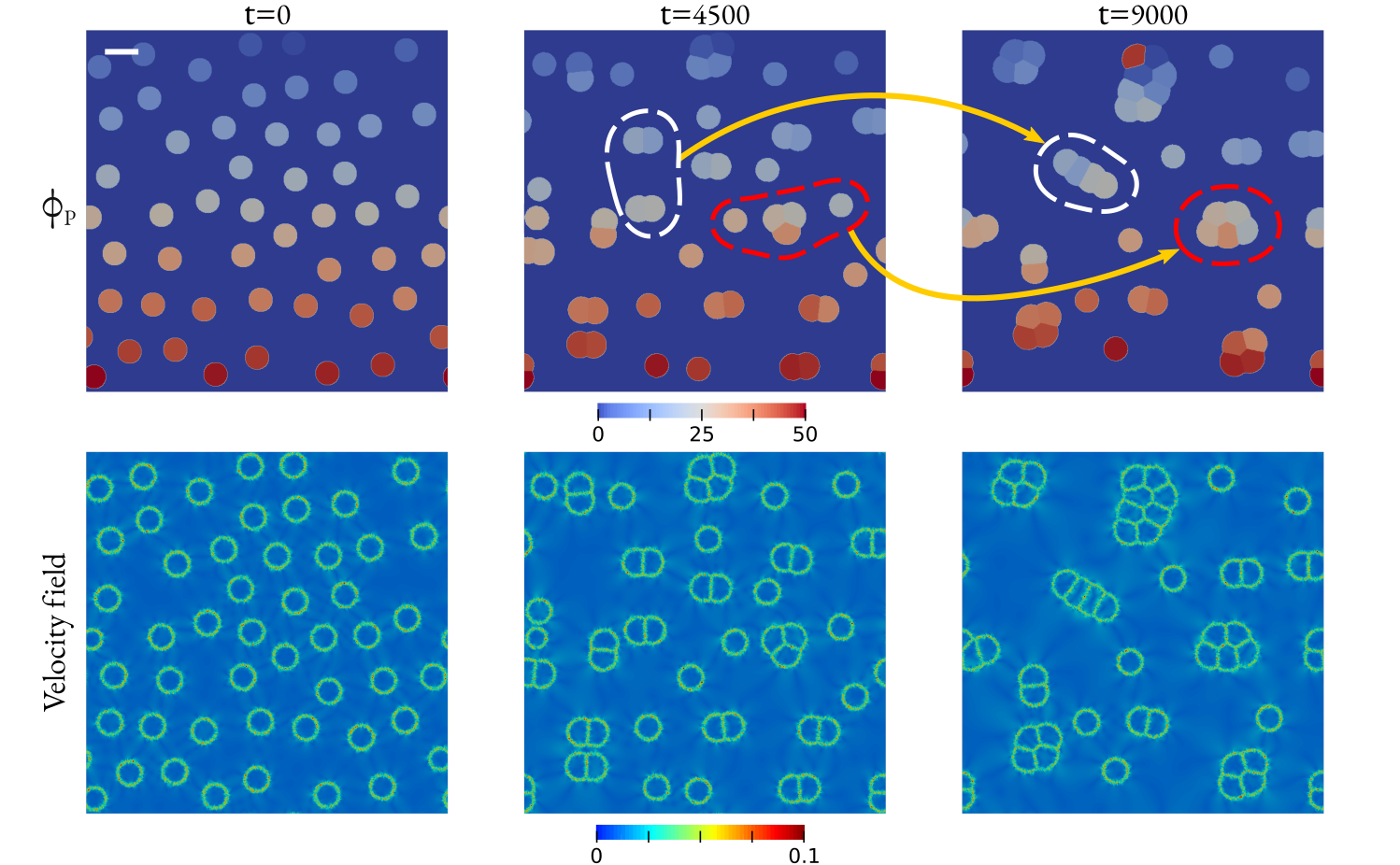}
 \centering
 \caption{The Brownian motion of multiple droplets propelled by the random composition noise fluctuation $\xi_{\scriptscriptstyle \phi}=0.5$ for all components in the ACNS model. 
 Upper row: 50 droplets coalescence with time. 
 Two typical morphology transformations: (i) rod formed by Janus particles, and (ii) big cluster via particle collisions are highlighted with the white and red dashed curves, respectively.
The underlying color bar indicates the $\boldsymbol{\phi}$ index of the system, 
with the first phase field variable $\phi=0$ for the matrix and the rest numerating the $50$ droplets. 
Lower row: velocity field around droplets scaled by the color bar beneath. 
The scale bar denotes the spatial length of $50$.
 }
 \label{fig:mp}
\end{figure*}

\textcolor{black}{Here, we report a coalescence mechanism distinct from the Golovin-Tanaka theory by applying the ACNS model. 
Since the obstacle potential is adopted to the system as Eq.~\eqref{eq:omega}, 
the composition profiles are depicted by sinusoidal functions 
with finite interfacial width. 
Therefore, in contrast to the hyperbolic tangent interface profile in the CHNS model,
the ACNS model enables us to eliminate the composition overlapping in the Golovin-Tanaka mechanism, 
and deletes the non-uniform composition induced Marangoni effect. 
For the simulation setup, 
we position two droplets of equal radius $r_{\scriptscriptstyle 0}=20$ 
symmetrically within a $N_{\scriptscriptstyle x}\times N_{\scriptscriptstyle y}=18r\times12r$ domain. 
The compositional noise with an amplitude $\xi_{\scriptscriptstyle \phi}=0.01$ is adopted. 
All other parameters are identical to that in Sec.~\ref{sec:acns_model}.
The initial separation between the particles 
is represented by the parameter $d$. }

\textcolor{black}{As shown in the double droplet simulations of Fig.~\ref{fig:2p}(I), 
the phase-field variable inside the gap is uniformly distributed as $\phi_{\scriptscriptstyle M}=0$. 
It implies that the surface tension force $-\phi\nabla\mu$ is strictly $0$ inside the gap region.
In other words, the composition noise induced Marangoni effect only appears at the interface region. 
But the fluid flows from each interface propagate into the matrix and overlap at the droplet gap region. 
Clearly demonstrated in Fig.~\ref{fig:2p}(II), 
the resulting flow velocity inside the gap region becomes larger than the outside part. 
Consequently, the non-uniform pressure distribution around droplets 
is established due to the composition noises (see Fig.~\ref{fig:2p}(III)) and produces the resulting force $-\nabla p$ which propels the droplet motion. }

\textcolor{black}{Most importantly, the composition noise induced droplet motion in dual-droplet setup shows three significant contrasts to the single droplet simulations.
(I) The droplet motion is intensified with the narrowing gap distance $d$.
By reducing $d$, 
the probability density function (PDF) of the droplet velocity
shifts towards higher velocities in Fig.~\ref{fig:2p}(IV).
(II) The droplet motion is anisotropic.
Fitting the displacements of 
the left droplet in the $x$ and $y$ directions 
independently to normal distributions, 
the $x$-displacement presents the apparent right shift with the reduction in $d$; see the middle panel of Fig.~\ref{fig:2p}(IV).
Contrarily, no discernible change related to droplet spacing is observed in the $y$ direction.
(III) The droplets motion is deterministic which can be 
proven by two aspects. 
(a) The $x$-displacement distribution shows an increase in the mean average value  with reducing $d$ in Fig.~\ref{fig:2p}(IV), 
indicating the deterministic drift. 
(b) The trajectories of the left droplet in Fig.~\ref{fig:2p}(V) apparently present the right shifting motion.}

\textcolor{black}{Particularly, 
as the asymmetric velocity field results in the pressure gradient $-\nabla p$ which points inwards the gap, 
the deterministic droplet motions always lead to the droplet coalescence. 
Hence, the composition noise induced coalescence force 
is pure attractive and independent of the surface tension gradient. 
It manifests the distinction between our observation and
the Golovin-Tanaka theory which indicates that the droplet coalescence always evolves to reduce the system's total surface energy. 
In other words, 
the coalescence of hydrophobic droplets cannot be explained with Golovin-Tanaka mechanism, 
but is indeed captured in our simulations; 
see Supplementary Sec. III, 
as well as the latest experiment~\cite{davoodianidalik2022fluctuation}.
Moreover, the coalescence force has a long-range feature, 
differing from the mechanism due to the short-range bridging effect at droplet reported in Ref.~\cite{perumanath2019droplet}.
An obvious proof is evident in Fig.~\ref{fig:2p}(V), 
when the droplet spacing $d=60$ is $3$ times of their radii, 
the deterministic merging can be apparently seen in the droplet trajectory.}

Furthermore, we proceed with the multi-particle simulation as demonstrated in Fig.~\ref{fig:mp}. 
The initial 50 particles with radii $r_{\scriptscriptstyle 0}=20$ are randomly distributed in a 2-dimensional domain ($N_{\scriptscriptstyle x}=N_{\scriptscriptstyle y}=30\,r_{\scriptscriptstyle 0}$). 
Propelled by the composition noise with the amplitude $\xi_{\scriptscriptstyle \phi}=0.5$ for all components, 
the particle motion results in a manifest coalescence behavior, 
as depicted in the first row of Fig.~\ref{fig:mp}. 
As a result of the asymmetric velocity field around the particles, 
distinct multi-particle microstructures are observed, 
including Janus, rod, and cluster, 
as highlighted by the colored dashed curves in Fig.~\ref{fig:mp}. 

\section{Conclusion}
In conclusion, we have postulated and validated two types of stochastic phase-field models coupling with hydrodynamics to simulate the Brownian motion of particles and droplets.
Propelled by the composition fluctuations with weak Marangoni effect, 
the particle/droplet proceeds the Brownian motion, 
depending on the amplitude of the random noise and the particle size, and the microscopic kinetic parameter. 
Moreover, by altering the parameters in the Navier-Stokes equations, 
the stochastic phase-field models go
beyond the limitation of the Langevin equation only for the rigid body, 
and can also be utilized for soft deformable droplets. 
After testifying our results with the Einstein relationship within the equilibrium condition, 
we extend our model further into two off-equilibrium scenarios.
i) When the composition noise-induced fluid flow becomes pronounced, 
the transition from Brownian motion to ballistic motion is observed which indicates 
the noise-induced fluid flow underdamped by the viscous stress.
ii) The double particle simulation shows 
a stochastic-induced deterministic droplet motion, 
which plays a vital role in the coalescence of the multi-particle system 
and is hardly considered in the Langevin dynamics. 
Nevertheless, in the previous Cahn-Hilliard type phase-field models~\cite{golovin1995spontaneous,tanaka1996coarsening}, 
the stochastic noise terms are totally overwhelmed by the pronounced diffusion 
and Ostwald ripening, 
and simply applied as a trigger for the phase separation. 
Here, we reevaluate the importance of the noise term 
and focus on the stochastic droplet motions during the coalescence process, 
while the subordinate Ostwald ripening effect can be neglected. 
Thus, the missing linkage between the coarsening droplet 
and the randomly drifting rigid particle is connected.
Our model is also fully implemented in 3D. 
We anticipate performing large-scale 3-dimensional multi-droplet simulations 
in forthcoming research to understand 
the underlying mechanisms of the micro-droplet motions, 
especially for the gelation process of soft matter materials. 

\begin{acknowledgments}
H.Z. thanks the Gottfried-Wilhelm Leibniz prize NE 822/31-1 of the German Research Foundation (DFG) for funding this research. 
F.W. is grateful to the VirtMat project P09  
of the Helmholtz Association (MSE-programme No. 43.31.01). 
The authors acknowledge support from the state of Baden-Wuerttemberg through bwHPC.
\end{acknowledgments}

\bibliography{aps}

\end{document}